\documentclass[pra,twocolumn]{revtex4-2}
\pdfoutput=1
\usepackage{epsfig}
\usepackage{graphicx}
\usepackage{amsmath}
\usepackage{natbib}
\usepackage{color}

\newcommand{\n}{\nonumber}
\newcommand{\bn}{\begin{eqnarray}}
\newcommand{\en}{\end{eqnarray}}
\newcommand{\eml}{\end{multline}}
\newcommand{\bml}{\begin{multline}}
\newcommand{\h}{\hspace}

\newcommand{\op}[1]{\hat{#1}}

\newcommand{\pt}{\partial_\theta}

\begin{document}

\title {Effects of a rotating periodic lattice on coherent quantum states in a ring topology: The case of positive nonlinearity}

 \author{Hongyi Huang $^1$ and Kunal K. Das$^{1,2}$}
 \affiliation{$^1$Department of Physics and Astronomy, Stony Brook University, New York 11794-3800, USA}
 \affiliation{$^2$Department of Physical Sciences, Kutztown University of Pennsylvania, Kutztown, Pennsylvania 19530, USA}
%
\begin{abstract}
We study the landscape of solutions of the coherent quantum states in a ring shaped lattice potential in the context of ultracold atoms with an effective positive nonlinearity induced by interatomic interactions. The exact analytical solutions in the absence of lattice are used as a starting point and the transformation of those solutions is mapped as the lattice is introduced and strengthened. This approach allows a simple classification of all the solutions into states with periods commensurate/incommensruate with the lattice period and those with/without nodes. Their origins are traced to the primary dispersion curve and the swallowtail branches of the lattice-free spectrum. The commensurate states tend to remain delocalized with increasing lattice depth, whereas the incommensurate ones may be localized. The symmetry and stability properties of the solutions are examined and correlated with branch energies. The crucial importance of rotation is highlighted by its utility in continuously transforming solutions and accessing in a finite ring with a few sites the full spectrum of nonlinear Bloch waves on an infinite lattice.
\end{abstract}

\maketitle

\section{Introduction}

A defining feature of Bose Einstein condensates (BEC) has been the effective nonlinear behavior induced by interatomic interactions, and a substantial literature exists on the subject \cite{Bloch-RMP-Many-Body}. The interplay of such interactions with a periodic lattice potential \cite{RMP-Oberthaler} has led to realizing significant phenomena such as the Mott insulator to superfluid transition \cite{Bloch-mott}. The ubiquity of harmonic trapping in experiments \cite{RMP-Sringari-1999} dictated that the bulk of the studies of ultracold atoms in periodic lattices have been in the context of open systems with trivial topology. In recent years, ultracold atoms have been trapped in ring configurations proving convenient for the study of superfluid features, like persistent flow and quantization of angular momentum \cite{ramanathan,Phillips_Campbell_superfluid_2013}. However, such efforts have not extended to include an azimuthal periodic lattice, although the capability exists \cite{Padgett, Zambrini:07}.

The non-trivial topology of a ring combined with the periodic lattice structure has already been shown to present a wealth of physical phenomena, whether examined with continuum \cite{Das-PRL-localization,Opatrny-Kolar-Das-rotation,Das-Brooks-Brattley, Aghamalyan-two-ring-lattice, Tiesinga-soliton-lattice,  Jezek-winding-number, Nigro_2018, Opatrny-Kolar-Das-LMG,Das-Christ} or discrete \cite{Doron-Cohen-1,Moreno-Bose-Hubbard,Jezek-Bose-Hubbard-ring-lattice,Maik-dipolar,Piza-ring-lattice,Minguzzi-PRA-two-bosons,Penna, Aghamalyan-AQUID, Minguzzi-resonant-persistent,Arwas-Cohen-PRB2017,Arwas-Cohen-PRA2019} lattice models. Adding nonlinearity via atomic interactions substantially increases the complexity of the system and hence the range of possible behavior. The system offers the possibility of studying nonlinear dynamics in a lattice system which is closed, finite and is naturally without boundaries. A necessary prelude to such a study is mapping out the space of allowed solutions and the effect of possible rotation.  That is the specific goal of this study.

The nonlinear states of a BEC in a lattice have been examined in several studies, experimentally \cite{Oberthaler-gap-soliton,Oberthaler-selftrapped} and theoretically \cite{RMP-solitons,Bronski, Wu_Niu,Konotop_Salerno,Pethick_Smith_PRA,smith03, smith04,Holland-Kronig-Penney,holland05}.  However, the primary focus of these studies have been systems with trivial topology and in the limit of large lattices. It is only recently that there has been a thorough study of nonlinear states focussed entirely on a continuum ring lattice configuration with a few periods \cite{Guilleumas-nonlinear_ring}. The approach taken there was to assume the presence of a lattice of fixed strength and examine the effect of introducing the nonlinearity. We take a different but complementary approach, where we examine the effects of introducing the lattice into the nonlinear system. There are multiple motivations for this. Prime among them is  that exact analytical solutions exist for such nonlinear systems, in the absence of a lattice \cite{carr00-a,carr00-b}. They provide concrete insights when the lattice is introduced with increasing strength. Secondly, in experiments adiabatic introduction of the lattice leads to some interesting dynamics in the linear regime as one of us recently showed \cite{Das-Brooks-Brattley}, and this current work provides the essential basis for a study of similar dynamics in the nonlinear regime. Finally, and significantly, this approach leads to a simpler and more transparent description of the spectrum and states, and their classifications and interconnections. This will be particularly relevant when examining the complex dynamics of such a system.

Our approach leads to several useful insights about coherent modes in ring lattices. The relation of the lattice period with that of the eigenmodes in the absence of the lattice determines the degree of localization of the modes when the lattice is present, and the origins are tied to the nature of the nonlinear spectrum in the absence of lattices. Gap solitons are found to be not a very distinctive feature for finite ring lattices. We identify certain symmetries of the lattice-free solutions, which persist even when the lattice is introduced and strengthened. Effects of the ring topology is particularly prominent for small lattice sizes, in the quantized modes. Increasing the number of lattice sites leads to emergence of proportionately more soliton branches. Rotation is an essential tool with ring lattices, allowing access to a continuous range of solutions, for a set lattice period and size.

In Sec.~II we set up our physical model, then identify the key effects of the ring boundary conditions and possible rotation in Sec.~III. We set the template for examining the states with an azimuthal lattice in Sec.~IV with a brief derivation and classification of nonlinear analytical solutions in a ring without a lattice. Section~V provides a comprehensive description of how the spectrum of nonlinear solutions transforms as the lattice is introduced and strengthened. The persistent symmetries of the eigenstates are identified in Sec.~VI and used to classify and characterize them, and effects of the lattice on the eigenstates are explained in Sec.~VII, specifically identifying why certain solutions remain delocalized and others tend to localize.  Section VIII compares our approach and results to those obtained by introducing nonlinearity with an existing lattice of fixed strength. We analyze the stability of the solutions by considering small fluctuations about the mean field solutions in Sec.~IX, examine the behavior in the limit of rings with large number of lattice sites in Sec.~X, and conclude in Sec.~XI with a brief summary and outlook for continuing work.

\section{Physical Model}
We consider a BEC in a toroidal trap similarly our prior work \cite{Das-Brooks-Brattley, Opatrny-Kolar-Das-rotation}. We take the minor radius to be much smaller than the major radius $R$ so that the system can be treated as a cylinder ${\bf r}=(z,r,\varphi)$ with periodic boundary condition on $z$.  We assume the confinement along $(r,\varphi)$, transverse to the ring circumference to be sufficiently strong to keep the atoms in the ground state
$\psi(r,\varphi)$ for those degrees of freedom, so that the three-dimensional bosonic field operator can be written in the effective form $ \hat{\Psi}(z)\psi(r,\varphi)$.
Integrating out the transverse degrees of freedom, the dynamics can be described by an effective one dimensional (1D) Hamiltonian
\begin{eqnarray}\label{QF-Hamiltonian}
\op{H}(t)&=&\int_0^{2\pi R}{\rm d}z\op{\Psi}^\dagger(z,t)\times\left[-\frac{\hbar}{2m}\partial_z^2+V(z,t)\right.\\
&&\left.+\frac{g_{3D}N}{4\pi l^2} \op{\Psi}^\dagger(z,t)\op{\Psi}(z,t)\right]\op{\Psi}(z,t).\n
\end{eqnarray}
where  $g_{3D}=4\pi\hbar^2a/m$ is the interaction strength defined by $a$ the $s$-wave scattering length ,  $m$ is the mass of individual atoms, $N$ is the total number of atoms, and $l=\sqrt{\hbar/m\omega_T}$ is the harmonic oscillator length for the transverse confinement along the cross-section of the torus. We assume a sinusoidal lattice potential $V(z,t)=V_0\sin^2\left(\frac{q}{2}(z/R-\Omega t)\right)$ rotating with angular velocity $\Omega$ with respect to the laboratory frame.

We take the major radius $R$ as the length unit so that the linear distance along the ring coincides with the angular distance $z/R=\theta\in [0,2\pi)$; we take the lowest energy scale in the ring $E_R=\frac{\hbar ^2}{mR^2}$ as the energy unit; we use the associated frequency $\omega_R=\frac{\hbar}{mR^2}$ as unit for frequency as well as the angular velocity, use it to set our time scale $\tau=\omega_R^{-1}$. The effective nonlinear constant in 1D then has the form $g=2a\omega_TN$, note we include the particle number in the constant.  Using these units leads to the equation of motion which, in the mean field limit $\langle\hat{\Psi}\rangle=\psi$, is a nonlinear Schr\"odinger equation:
\begin{equation}
\label{eq1.1-9}
\left[{\textstyle\frac{1}{2}}(i\partial_\theta+\Omega)^2+V_0\sin^2({\textstyle\frac{1}{2}}q\theta)+g|\psi|^2\right]\psi=i\partial_t\psi
\end{equation}
The explicit presence of $N$ implies the normalization $\int_0^{2\pi}d\theta \left|\psi(\theta,t)\right|^2 =1$.  The presence of the angular velocity $\Omega$ in above equation represents transformation to the frame rotating with the lattice, which remove explicit dependence on time in the lattice potential.  A centrifugal term $\propto \Omega^2$ is left out as being a constant offset for fixed radius $R$.  The stationary solutions $\varphi(\theta)=\psi(\theta,t)e^{i\mu t}$ satisfy the time-independent version of Eq.~(\ref{eq1.1-9}) with $i\partial_t\rightarrow \mu$ where the eigenvalues $\mu$ of the equation defines the chemical potential.  That equation can be considered the stationary solution of the energy functional
\bn
\label{energy_functional}
&&E\left[\varphi\right]=\int_0^{2\pi} d\theta\times\\
&&\left[{\textstyle\frac{1}{2}}\varphi^*(i\partial_\theta+\Omega)^2\varphi+V_0\sin^2({\textstyle\frac{1}{2}}q\theta)|\varphi|^2
+{\textstyle\frac{1}{2}}g\left|\varphi\right|^4\right]
\n\en
with the chemical potential as the Lagrange multiplier for the normalization $\int_0^{2\pi} d\theta|\psi|^2=1$ condition, and related to the mean field energy as $\mu=E\left[\varphi\right]+{\textstyle\frac{1}{2}}g\int_0^{2\pi}d\theta |\varphi|^4$.

\begin{figure}[t]
\centering
\includegraphics[width=\columnwidth]{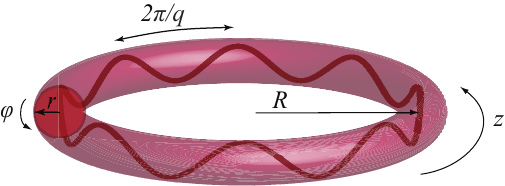}
\caption{(Color online) The atoms are trapped in a toroidal trap with an azimuthal lattice potential of period  $2\pi/q$, its variation of depth shown schematically as a thick sinusoidal line. The torus is taken as a wrapped cylinder with our choice of co-ordinates ${\bf r}=(z,r,\varphi)$ shown, assuming the major radius to be much larger than the minor radius, $R\gg r$.  }
\label{Fig1-nolattice}
\end{figure}

\section{Boundary Conditions and Rotation}
The behavior of the physical observables can be understood best by writing the mean field stationary state in the polar amplitude-angle form, referred to as the hydrodynamic picture, $\varphi(\theta)=\sqrt{\rho(\theta)}e^{i\phi(\theta)}$, leading to an equation for the density $\rho$
\begin{equation}
\label{eq1.2-2}
{\textstyle\frac{1}{8}}(\pt\rho)^2-{\textstyle\frac{1}{4}}\rho\pt^2\rho+{\textstyle\frac{1}{2}}\alpha^2+V_0\sin^2({\textstyle\frac{1}{2}}q\theta)\rho^2+g\rho^3-\mu\rho^2=0
\end{equation}
and a phase equation that provides an integral of motion,
\bn\textstyle
\rho\pt\phi-\Omega\rho&=&\alpha\\ \Delta \phi (\theta)=\phi (\theta)-\phi (0) &=&\Omega\theta+\int_{0}^{\theta}\frac{\alpha}{\rho(\theta')}d\theta'.
\label{eq1.2-1}
\n\en
This sets the current density $J=N\alpha$, the superfluid velocity $v=\alpha/\rho(\theta)$ and angular momentum per particle $L=\hbar\Omega+2\pi\hbar\alpha$.

The single-valuedness of the quantum wave function and the closed topology of the ring impose the following boundary conditions
\bn
\label{eq1.3-2}
\rho(0)=\rho(2\pi) \hspace{5mm}
\rho'(0)=\rho'(2\pi) \hspace{5mm}
\Delta \phi (2\pi)
=2\pi n,
\en
with the integer $n$ being the winding number. The total phase change contains the effect of rotation. We will use the \emph{bare} phase change around the ring neglecting rotation
\bn \delta\phi\equiv\delta\phi(2\pi)=\int_{0}^{2\pi}\frac{\alpha}{\rho(\theta')}d\theta'= \Delta\phi(2\pi)-2\pi\Omega.\label{bare_phase}\en

\begin{figure}[t]
\centering
\includegraphics[width=\columnwidth]{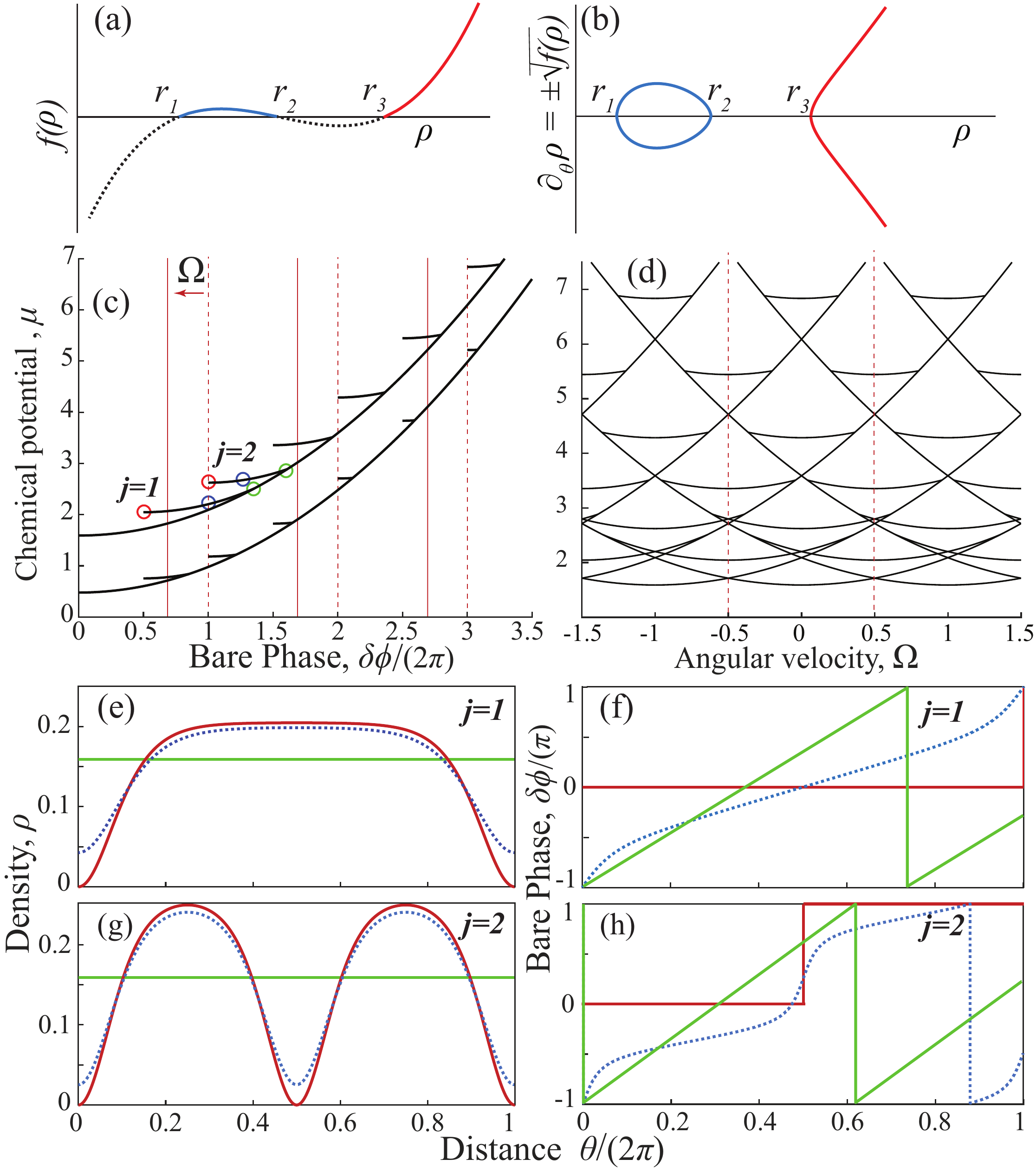}
\caption{(Color online)(a) Shape of the cubic function that sets the density variation, displaying its roots; the dotted parts are forbidden, (b) Loop in phase space that satisfies the boundary conditions on a ring, and corresponds to the solid sections in panel (a). (c) The mean field chemical potential $\mu$ in the absence of a lattice, $V_0=0$, for two interaction strengths $g=3$ and $g=10$; the soliton branches indexed by $j$ lengthen as nonlinearity increases. Quantization of the total phase around the ring only allows solutions marked by intersections with vertical dashed lines, which can be continuously shifted by rotation, by any value of the angular momentum $\Omega\neq 0$ indicated schematically by the solid lines. (d) The case of $g=10$ in periodic zone scheme as a function of rotation. Lowest panels display the eigenstates for $\mu$ marked by circles in (c): Their densities are shown in (e,g) with plane wave (flat green line) at the right edge of the branch, solutions with nodes (red line) at the left edge, and one example of solutions without nodes (dotted blue) that lie in between. Corresponding bare phase variations $\delta \phi$ are shown in (f,h).}
\label{Fig1-nolattice}
\end{figure}

We can understand the system behavior by tracking the chemical potential $\mu$ as function of the bare phase change $\delta\phi$. Insisting on the density boundary conditions, as we vary $\alpha$,  yields a continuum of solutions such as illustrated in Fig.~\ref{Fig1-nolattice}(c) for the case when the lattice is absent. However, the phase boundary condition for a finite size ring picks out only discrete points on that spectrum as physically relevant, marked on the plot by dashed vertical lines. Rotation shifts those vertical lines by adding a phase ramp, allowing access to the complete spectrum.  Owing to the complementary relation $2\pi\Omega=-\delta\phi+2\pi n$, the angular velocity $\Omega$ can replace the bare phase, in the representation of the spectrum as shown in a periodic zone scheme in Fig.~\ref{Fig1-nolattice}(d). This can compared to the limit of no potentials presented in a recent work \cite{Obiol-Cheon}.

\section{Exact solutions without lattice}
\label{sec2}
In the absence of a lattice $V_0=0$, solutions of Eq.~(\ref{eq1.1-9}) can be found analytically \cite{carr00-a,carr00-b} which will provide the framework for our description once the lattice is introduced. A first integration of Eq.~(\ref{eq1.2-2}) yields
\bn
\label{eq.hd.3}
\pt\rho=\pm\sqrt{f\left(\rho\right)}; \h{3mm}
f\left(\rho\right)=4g\rho^3-8\mu\rho^2+8\beta\rho-4\alpha^2
\en
with integral of motion $\beta$.  The nature of the solutions can be related to the root structure of the function $f(\rho)$ which has a general shape shown in Fig.~\ref{Fig1-nolattice}(a) for repulsive nonlinearity, $g>0$. It is clearly required that $f(\rho)\ge 0$, corresponding to the colored solid segments in the plot, and the roots of $f(\rho)= 0$, labelled $r_1\leq r_2\leq r_3$, need to be real to be physically relevant. Viewing $\rho'$ and $\rho$ as parametric functions of $\theta$ in a phase space representation, their joint periodic boundary conditions in Eq.~(\ref{eq1.3-2}) require that $\pm\sqrt{f}$ plotted as a function of $\rho$ forms one or more complete circuit around the loop drawn in Fig.~\ref{Fig1-nolattice}(b) as $\theta$ is varied from $[0,2\pi)$. With the cubic form for $f$ this is only possible if we combine $+\sqrt{f}$ and $-\sqrt{f}$ to form the two halves of the loop meeting at the roots $r_1$ and $r_2$.

The roots define the state of the system. To start with the symmetry of the loop assures that the state will always have at least two points of reflection symmetry along the ring, which will be relevant when the lattice is turned on as we will discuss in Sec.~\ref{Sec:symmetry}. The density varies between the roots $r_1$ and $r_2$ suggesting parametrization as $\rho=r_1+(r_2-r_1)t^2$. Choosing $\rho(\theta=0)=r_1$,  Eq.~(\ref{eq.hd.3}) can be integrated as
\bn\label{density_theta}{\textstyle
\theta=\frac{1}{\sqrt{g(r_3-r_1)}}}\int_{0}{\textstyle^{\sqrt{\frac{\rho(\theta)-r_1}{r_2-r_1}}}\frac{dt}{\sqrt{(1-t^2)(1-mt^2)}}}
\en
with the relevant parameters also in terms of the roots
\bn\label{parameters}\textstyle
m&=&\frac{r_2-r_1}{r_3-r_1},\h{5mm} \mu=\frac{g}{2}(r_1+r_2+r_3)\n\\
\alpha^2&=& gr_1 r_2 r_3,\h{5mm}
\beta=\frac{g}{2}(r_1 r_2+r_1 r_3+r_2 r_3).
\en
Inversion of Eq.~(\ref{density_theta}) yields a Jacobi elliptic function \cite{abramowitz1964handbook}, and using it in Eq.~(\ref{bare_phase}) determines the phase in terms of an incomplete elliptic integral of the third kind \cite{byrd13book},
\bn
\rho(\theta)&=& r_1+(r_2-r_1){\rm sn} ^2(\textstyle{\sqrt{g(r_3-r_1)}}\ \theta,m) \n\\
\delta\phi(\theta)&=&\frac{\alpha}{r_1\sqrt{g(r_3-r_1)}}\Pi(1-\frac{r_2}{r_1},\varphi,m)
\en
where $\varphi=\sin^{-1}[{\rm sn}(\sqrt{g(r_3-r_1)}\ \theta,m)]$.
Taking the smallest value of $\theta\geq 0$ such $\rho(\theta)=r_2$, Eq.~(\ref{density_theta}) yields $\theta=K(m)/\sqrt{g(r_3-r_1)}$ which therefore sets the period of any density modulation. This however corresponds to only half of the loop in phase space. To satisfy the density boundary condition, the \emph{complete} phase space loop in Fig.~\ref{Fig1-nolattice}(b) has to be traversed by an integer number of turns we denote by $j$, leading to the condition for a complete circuit of the ring,
\begin{equation}
\label{eq.hd.8}
jK(m)=\pi\sqrt{g(r_3-r_1)}.
\end{equation}
The bare phase change around the ring is given by complete elliptic integrals of the first and the third kinds, $K(m)$ and $\Pi(m)$
\bn \delta\phi=\delta\phi(2\pi)&=&\frac{2\pi \alpha}{K(m) r_1}\Pi(1-\frac{r_2}{r_1},m).
\en

\begin{figure*}[t]
\centering
\includegraphics[width=\textwidth]{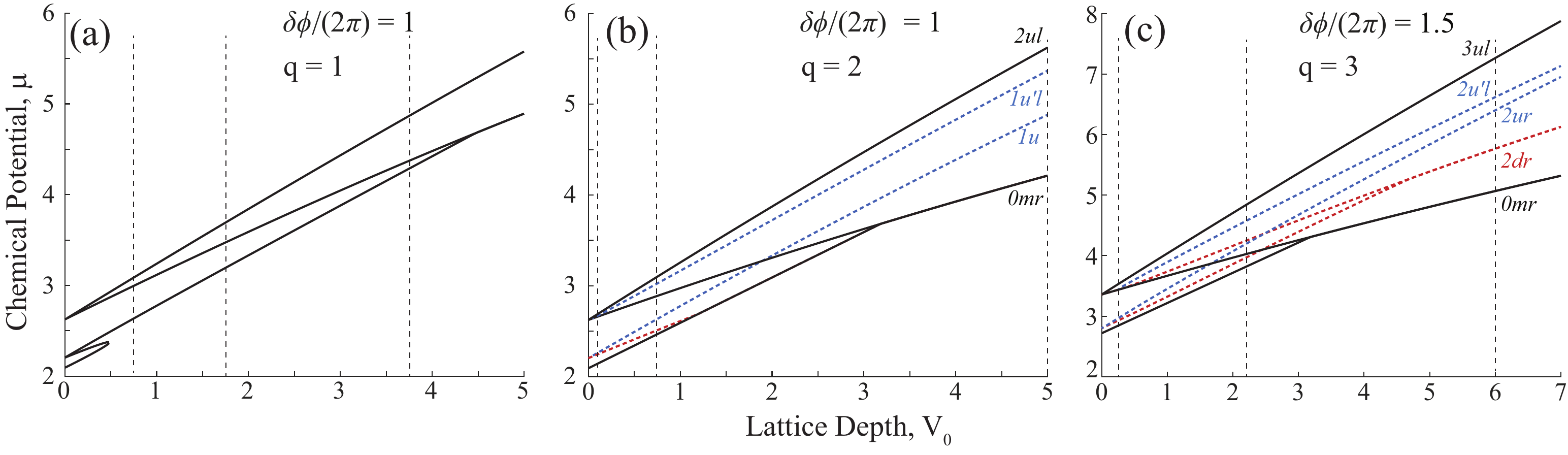}
\caption{(Color online) The chemical potential $\mu$ plotted as a function of the lattice depth $V_0$ for the number of lattice periods $q=1,2,$ and $3$ in panels (a,b,c). The nonlinear strength is fixed at $g=10$. The specific bare phase values accumulated around the ring $\delta\phi$ are indicated in each panel and correspond to the vertical dotted lines in Fig.~\ref{Fig3-nonlinear-spectrum}.  The vertical dashed lines that appear here in turn indicate the values of $V_0$ that correspond to the plots of $\mu$ versus $\delta\phi$ that appear in Fig.~\ref{Fig3-nonlinear-spectrum}.  The labels for the different branches follow the convention defined in Eq.~(\ref{convention}) also in the context of the next figure.  The dashed lines correspond to the intra-band soliton branches in the next figure, with blue and red marking upper and lower ones respectively of the split branches. }
\label{Fig2-VaryV0}
\end{figure*}

\begin{figure*}[t]
\centering
\includegraphics[width=\textwidth]{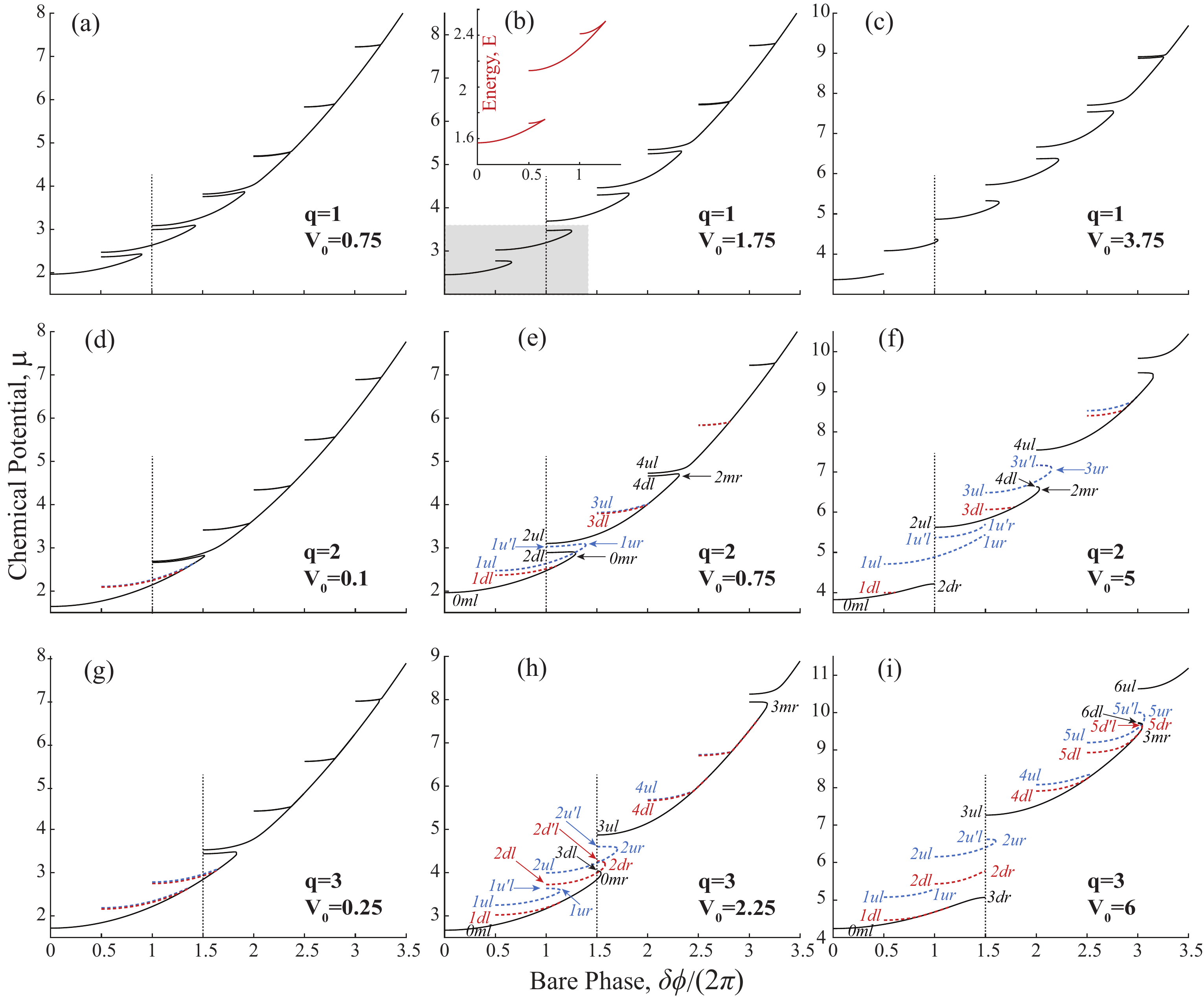}
\caption{(Color online) The chemical potential $\mu$ plotted as a function of the phase $\delta\phi$ for three different lattice periods $q=1,2,$ and $3$, one in each row.  The nonlinear strength is fixed at $g=10$. The three panels in each row are for different values of the lattice depth $V_0$ corresponding to the vertical dotted lines in Fig.~\ref{Fig2-VaryV0}. The vertical dotted lines here indicate the phase values $\delta\phi$ that correspond to the plots in Fig.~\ref{Fig2-VaryV0}. The inset in panel (b) shows that the rounded hook structures appear as swallowtails when total energy rather than chemical potential is plotted, here illustrated for the gray highlighted region.  The labels for the different branches follow the convention defined in Eq.~(\ref{convention}). The intra-band soliton branches are shown in dashed lines, and when they split, the upper branches are shown in blue and the lower ones in red.}
\label{Fig3-nonlinear-spectrum}
\end{figure*}

The normalization of the density provides an additional constraint involving the complete elliptic integral of the first and second kinds $K(m)$ and $E(m)$
\bn\label{norm} \textstyle \int_{0}^{2\pi}\h{-2mm}\rho d\theta=2\pi r_1+2\pi(r_3-r_1)[1-E(m)/K(m)]=1. \en
The definition of $m$ in Eq.~(\ref{parameters}) along with Eqs.~(\ref{eq.hd.8}) and (\ref{norm}) can be used to express the roots in terms of the complete elliptic integrals of the first and second kinds
\bn r_1&=&{\textstyle \frac{1}{2\pi}+(E-K)\frac{j^2K}{\pi^2g}}\n\\
r_2&=&{\textstyle \frac{1}{2\pi}+(E-K+mK)\frac{j^2K}{\pi^2g}}\n\\
r_3&=&{\textstyle \frac{1}{2\pi}+\frac{j^2 EK}{\pi^2g}}.\en
When $j$ and $g$ are specified, these are completely determined by the value of $m$, which can be determined by imposing the phase boundary condition, yielding the complete solution.

The resulting spectrum is illustrated in Fig.~\ref{Fig1-nolattice}(c) comprising of a parabolic dispersion curve and a sequence of branches given the fanciful name of swallowtails due to their appearance in the periodic zone scheme \ref{Fig1-nolattice}(d). The spectrum represents three different behaviors as discussed below. Examples for $j=1$ and $j=2$ are shown in Fig.~\ref{Fig1-nolattice}(e-h).

{\bf Plane waves}:  Spectral values on the parabolic dispersion curve correspond to plane wave solutions as would be the case in the linear regime. The solutions arise when two or more of the roots of $f(\rho)$ are real and degenerate, and both $f(\rho)=0$ and $f'(\rho)=0$ in the phase space plot. The density being uniform, these are without nodes
\bn
\label{eq2.2-1}{\textstyle
\psi(\theta)=\sqrt{\frac{1}{2\pi}}e^{in\theta},\ \ \alpha=\frac{n}{2\pi},\ \
\mu=\frac{n^2}{2}+\frac{g}{2\pi}}
\en
with $n=0, \pm1, \pm2, \cdots$.  These appear with uniform density and a phase ramp in Fig.~\ref{Fig1-nolattice}(e-h).

{\bf Solutions with nodes}: In the extended zone scheme, the swallowtail branches terminate at multiples of the Brillouin zone (multiples of $\pm \pi$ in our units).  These termination points correspond to solutions with nodes, and the stationary states can be taken to be real valued everywhere, which implies as we approach these nodes, $\alpha\rightarrow 0$ and the bare phase $\delta\phi\rightarrow \pm j\pi$ so that at each node there is a phase slip of $\pm\pi$ \cite{Carr_Ueda_PRL}. The states and the corresponding chemical potentials are limiting cases of the general solution above when $r_1 \rightarrow 0$
\begin{equation}
\label{eq2.1-1}
\begin{array}{c}
\psi(\theta)=\sqrt{\frac{m}{2\pi\left(1-E/K\right)}}{\rm sn}\left(\frac{jK\theta}{\pi}| m\right),\
\mu=\frac{(1+m)j^2K^2}{2\pi^2}
\end{array}
\end{equation}
The index $j$ enumerates the nodes of the wave function. Since the phase can jump at the nodes, in the absence of rotation, the periodic boundary condition requires even integers $j=2, 4, 6, ... $, but odd values can be accessed by rotation.   Eqs.~(\ref{eq.hd.8}) and (\ref{norm}) lead to the condition
\begin{equation}
\label{eq2.1-2}
\frac{\pi g}{2j^2}=K(m)^2-K(m)E(m).
\end{equation}
Examples of such solutions are shown in Fig.~\ref{Fig1-nolattice}(e-h), in solid curves with density vanishing at $j$ points, with phase jumps of $\pi$ occuring at those points. The nodes occur at the tips of the swallowtails which correspond to extrema of the chemical potential with respect to both $\Omega$ and $\delta \phi$, due to the relation between them, and evident as symmetry points in the periodic zone view Fig.~\ref{Fig1-nolattice}(d).  Assuming the conditions for nodes $\rho=\rho'=0$, it can be easily seen that $\partial E/\partial\Omega=0$ at those points, so the energy has extrema, as well as, $\mu=\partial E/\partial N$. These continue to be true with the lattice on.

{\bf Nodeless Density Modulations}: The remaining stretch of each swallowtail branch is described by the general solution defined above, and correspond to nodeless complex-valued states with density modulation with $j$ dips. These are intermediate between the nodeless plane waves at one end of the branch and the solutions with nodes at the other end. The bare phase change is a monotonically \emph{decreasing} function of the elliptic parameter $m$, and for each specific branch labelled by $j$, the phase lies in the range $[j\pi,\sqrt{j^2\pi^2+2\pi g}]$ corresponding to $m\in [0,m_c]$. The upper limit of the elliptic parameter $m_c$ set by Eq.~(\ref{eq2.1-2}) marks the solutions with nodes and the lower limit $m=0$ corresponds to plane waves. In Fig.~\ref{Fig1-nolattice}(e-h), we see examples plotted in dotted lines; the density is periodically modulated, but it never vanishes, and the bare phase has a more complicated position dependence.

\section{Spectrum with Lattice}

We now examine how the nonlinear spectrum in the ring, as discussed in the last section, is impacted by the introduction of the lattice potential. In general, this needs numerical determination of the stationary solutions of Eq.~(\ref{eq1.1-9}). We do a Fourier expansion of the wave function in a plane wave basis $\psi=\frac{1}{\sqrt{2\pi}}\sum_{n=-\infty}^{\infty} c_n e^{in\theta}$ which leads to a set of coupled nonlinear equations in the momentum space amplitudes \cite{Das-Brooks-Brattley}
\bn
&&{\textstyle\frac{1}{2}}\{(n-\Omega)^2+V_0-2\mu\}c_n-{\textstyle\frac{1}{4}}V_0c_{n-q}-{\textstyle\frac{1}{4}}V_0c_{n+q}
\n\\
&&\h{2cm}+g\sum\limits_{j,k}c_k^{*}c_lc_{n+k-l}=0.
	\label{GPE_PW}\en
They are solved iteratively by Newton's method, until convergence criteria are met. In practice, only a finite number of modes are needed, typically between $10$ and $20$, with larger values of $g$ and $V_0$ requiring more.

As the lattice potential is turned on, Fig.~\ref{Fig2-VaryV0} shows that the chemical potential splits and at higher values of the lattice potential,$V_0$ new degeneracies can also appear as some of the split branches come together again. This occurs for any value of latticed period, as seen for the three values of $q=1,2,3$. Notably due to the ring configuration, even one lattice site $q=1$ constitutes a periodic lattice. However, more insight can be gained by plotting the chemical potential as a function of the bare phase at different strengths of the lattice potential, shown in Fig.~\ref{Fig3-nonlinear-spectrum} for specific values of $V_0$ that are marked by vertical dashed lines in Fig.~\ref{Fig2-VaryV0}.

As for a linear system, turning on the lattice has the usual effect of opening up gaps in the spectrum. But, in addition, when the lattice is still weak, the swallowtail branches split as well, as seen in Fig.~\ref{Fig3-nonlinear-spectrum}(a,d,g). Two distinct behavior emerge:  At the band edge, such splitting and the band gaps create hook like structures that terminate at the band edge.  The intra-band swallowtails, shown as colored dashed lines in that figure, each simply splits into two with the ends still terminating at the main dispersion curve, shown as solid black lines, which marked the plane wave solutions in the absence of the lattice, and transform to nonlinear Bloch wave solutions with the lattice on. The special case of $q=1$ obviously has no intraband swallowtails.

As the lattice depth increases, several things happen. The band gaps for the main branches widen as can be expected even in the linear case, but additionally the hooks shorten and can eventually vanish, as the lattice depth overwhelms the effect of the nonlinearity \cite{Wu_Niu-landau-zener}. The splitting of the intra-band swallowtails widen as well as can be seen in Fig.~\ref{Fig3-nonlinear-spectrum}(d-f), (g-i). Some of them eventually separate from the main branch and inherit existing hook structures in the main band (panels (e) and (h)), and they then follow the same shrinking trend with increasing lattice depth. As the band gaps widen, in panels (e,f,h,i) we see that some of the detached intra-band swallowtails end up in the band gap. They can be thought to correspond to what are called gap solitons \cite{RMP-solitons,Oberthaler-gap-soliton,louis03} in the case of large open lattices.

The structure of the analytical solutions without lattice allows us to classify the solutions with the lattice on. We use a three part label
\bn [j\mbox{-}index]\ [subbranch]\ [location],\h{5mm} j=0,1,2,\cdots\n\\
subbranch= \{m,u/u',d/d'\}, \h{5mm} location=\{l,r\}\label{convention}\en
with $j$ being the swallowtail branch index based on the analytical solutions, with $0$ representing the ground band section of the main dispersion curve once gaps open; the \emph{sub-branch} labels are $m$ for main branch, $u,d$ for up/down marking upper/lower branches of the split swallowtails and $u',d'$ for the hook section of detached intra-band swallowtails; and the \emph{location} labels $l,r$ for left or right end of each branch terminating at some multiple of $\pi$. Note,when a hook disappears, the location label can switch, as for example $2dl\rightarrow 2dr$ in Fig.~\ref{Fig3-nonlinear-spectrum}(e)$\rightarrow$(f) and $3dl\rightarrow 3dr$ in Fig.~\ref{Fig3-nonlinear-spectrum}(h)$\rightarrow$ (i).

Our representation provides a different and more unified perspective on the various solutions and how they are related. Exclusive focus on the spectrum and states in the absence of rotation is limited to a zero measure subset of the full range of solutions and hides the relations between various solutions. For example in Fig.~\ref{Fig3-nonlinear-spectrum}(h), in the absence of rotation, the lowest energy eigenstates would include the Bloch wave state $0ml$ at  $\delta\phi/(2\pi)=0$ along with states that are on the
vertical slice at $\delta\phi/(2\pi)=1$, which in our notation, includes the soliton solutions with nodes, $1u'l,2dl, 2ul$ and solutions without nodes lying on the $1d,1u$ branches, along with a nonlinear Bloch wave solution on the main branch. Of course, by symmetry the counterparts at $\delta\phi/(2\pi)=-1$ would be included. But, in our representation, we can also see that all the soliton solutions $1dl,1ul,1u'l$ originate from the splitting of the $j=1$ swallowtail branch and $2dl, 2ul$ from the splitting of $j=2$ branch in the absence of the lattice; and that $1ul,1u'l$ have a continuous range of nodeless states connecting them, all accessible by rotation.

\begin{figure*}[t]
	\centering
	\includegraphics[width=1\textwidth]{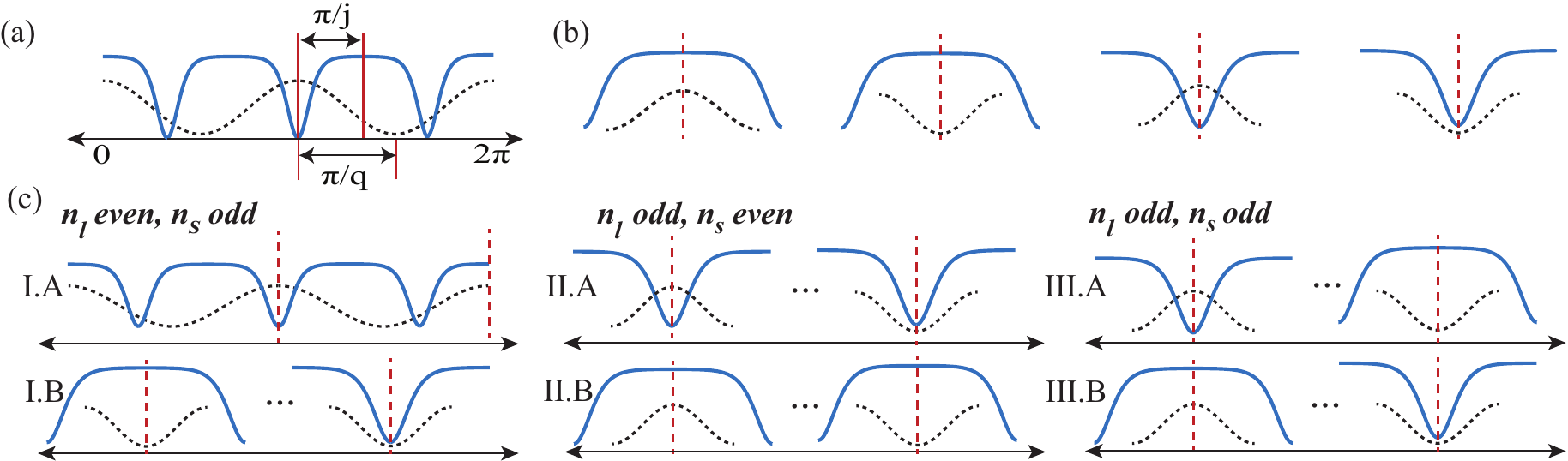}
	\caption{(a) The lattice period (dashed red line) does not typically match the period of the density modulation of the eigenstates (solid blue line) before the lattice is introduced. (b) When the lattice is barely introduced, the allowed eigenstates adapt to match the lattice period so that a subset of their separate maxima and minima line up at certain symmetry points in the ways shown. (c) There are three distinct ways how maximum/minimum of the lattice and of the eigenstates \emph{without} the lattice line up at \emph{adjacent} symmetry points, as the lattice is barely introduced. Each case has two subcases with \emph{Type-A} always containing a minimum of the density lining up with a maximum of the lattice.}
	\label{fig4-symmetry}
\end{figure*}

The full spectrum as displayed here also provides an alternate perspective on gap solitons as arising from intra-band swallowtail branches $j\neq nq$ (for integer $n$) as the lattice opens gaps among states on the main band. The intra-band swallowtails can detach from the main branch and end up in the gap, for example $1u'l$ in plot Fig.~\ref{Fig3-nonlinear-spectrum}(e) and $1ur, 2dl, 2ul$ in plot (i).  We can also see clearly the sensitivity of their appearance on the lattice depth. As the lattice depth increases, new gap solitons can appear with different properties, for example, progressing from plot (e) to (f), a new nodeless solution on the branch $1ul-1ur$ satisfies the boundary condition in the first gap.  The widening gap also can accommodate more solitons, for example in panel (i) has $1ul, 1ur$ that were not in the gap in (h).

The above statements need to be taken with the caveat that gap solitions are less well defined in a small ring lattice of few sites, inasmuch as, for any fixed angular velocity (including $\Omega=0$) the bands themselves are not continuum and comprise of $q$ discrete states allowing for positive and negative quasi-momenta, so the distinction between states in the `bands' and states in the `band gaps' is not as well demarcated . Here, we designate as gap solitons any state with its energy lying in the gaps of the main dispersion curve (solid black lines in Fig.~\ref{Fig3-nonlinear-spectrum}) for specific nonlinearity and lattice depth.

Rotation can clearly affect the set of allowed solutions dramatically in a ring lattice: Imagine sliding a set of vertical lines at $\delta\phi/(2\pi)=1,2,3,\cdots$ on any of the plots in Fig.~\ref{Fig3-nonlinear-spectrum} and sliding them together continuously, and wherever they intersect the spectrum are allowed solutions. Thus gap solitons can emerge and disappear, and states with nodes (at the tips of the branches) morph into nodeless solutions and eventually into a solution with different number of nodes, for instance in plot (i), transitioning a solution with two nodes at $1ur$ for $\Omega=0$ to a solution with one node at $1ul$ for $\Omega=\pi$ with a continuum of nodeless solutions in between.

Our representation also highlights certain interesting  features that were not sufficiently stressed in earlier studies. An intraband hook structure can actually split into two separate branches, creating two distinct solutions, as seen with $1ur$ in plot (e) splitting into $1u'r$ and $1ur$ in plot (f). Also, it is evident from their swallowtail origins that soliton branches can have the tips at any integer and half-integer multiples of $\delta\phi/(2\pi)$ and not just at the edge of Brillouin zone as sometimes implied.

\section{Symmetry of Eigenstates}\label{Sec:symmetry}

It is obvious that the lattice, as well as the modulated eigenstates in the \emph{absence} of the lattice, such as plotted in Fig.~\ref{Fig1-nolattice}, have reflection symmetry about their respective maxima and minima. The index $j$ enumerates the density nodes in the allowed eigenstates in the absence of the lattice, and in the rest of the branch these nodes transform into uniformly spaced density dips that get shallower and eventually vanish on merging into the main branch that corresponds to plane waves. The modulatated solutions without a lattice are a manifestation of spontaneously broken symmetry \cite{carr09}. When the lattice is turned on, the translational invariance is broken up to a lattice period, and the eigenstates adapt to have their points of reflection symmetry line up with those of the lattice, which we will call symmetry points. Notably, even a weak lattice is sufficient to cause this physical realignment, we have observed this for lattice strength as low as $V_0=0.01$ in our units, relative to kinetic energy $\sim 1$ and nonlinearity $g=10$.

This indicates four possible pairings of lattice max/min with the density max/min.  We find there is a symmetry of the density modulation that remains invariant when the lattice is turned on and even in the presence of rotation, as we now describe. Consider the smallest pair of mutually prime integers, $n_s$ and $n_l$ that enumerate the half periods of an eigenstate and of the lattice respectively that separate adjacent symmetry points, so that
\begin{equation}
\label{symmetry_condition}
\frac{\pi}{j}n_s=\frac{\pi}{q}n_l
\end{equation}
Even values for either integer would indicate repetition of maximum or minimum at every symmetry point and odd values indicate alternating of maximum and minimum. The case of both being even is clearly left out. That leaves us with three possibilities for when the adjacent symmetry points are different: (I) $n_l$ is even and $n_l$ is odd (II) $n_s$ is even and $n_l$ is odd (III) $n_s$ is odd and $n_l$ is odd.  Each of these has two sub-cases: A: when a lattice maximum and a state minimum coincide at one of the symmetry points and B: when a lattice maximum and state maximum never coincide at any of the symmetry points. All of these cases are illustrated in Fig.~\ref{fig4-symmetry}.

As the lattice is turned on, and even if there is rotation to access to all points on a branch, the number of dips and crests in the density can change, and even local minima and maxima can switch.  However, the key point is that the location of the symmetry points remain unchanged, meaning that those points continue to have reflection symmetry, even in the presence of the lattice and rotation.

We illustrate this in Figs.~\ref{Fig5-statesM2} and \ref{Fig6-statesM3} with eigenstates for lattice with $q=2$ and $q=3$ sites corresponding to spectra shown in Fig.~\ref{Fig3-nonlinear-spectrum}.  It is clear that where the branch index is a multiple of the lattice period $j=n\times q$, band gaps open up and the mutual commensurateness leads to nonlinear Bloch waves, as seen for panels (a) and (b) in both figures. The remaining panels show instances of the different symmetry cases mentioned above and are labelled as such. The states without lattice are shown as filled gray shapes and the lattice itself in long dashed lines, and the symmetry points lie at the \emph{center} and the \emph{edges} of each panel. When each soliton branch splits with the introduction of the lattice, we observe that the lower branch (labelled by second index `d') belong to Type-A, where at least one of the density minimum coincides with a lattice maximum; this is seen in the left panels (c) and (g) in Fig.~\ref{Fig5-statesM2} and the left panels (c,e,g,i) in Fig.~\ref{Fig6-statesM3}. Likewise, we observe the upper branch (labelled by second index `u') follows Type-B as seen in the right panels of both figures.

It is consistent energetically that the lower energy branches correspond to Type A, where a density minimum is at the lattice maximum. Notably this is the case for the Bloch wave states as well: At the band gaps which also appear as splitting of swallowtail branches with $j=n\times q$, each lower branch, corresponding to the top of the lower band, is of Type A, with density minima matching lattice maximum; and each upper branch corresponding to the bottom of the upper band is of Type B, which in the case of Bloch waves also ensure that density maxima match lattice maxima. This can be seen with states $2dr$ and $2ul$ in Fig.~\ref{Fig5-statesM2}(a,b) and states $3dr$ and $3ul$ in Fig.~\ref{Fig6-statesM3}(a,b).

In Fig.~\ref{Fig5-statesM2} we also illustrate the evolution of the states with increasing lattice depth, first at $V_0=0.75$ in panel (d) and then at $V_0=5$ in panels (e) and (f).  The symmetry points remain fixed, even though the shape of the states change substantially and even as new states with nodes $1ur$ and $1u'r$ emerge as the hook structure of upper branch breaks at the right edge, as noted at the end of the previous section.

We conclude this section with comments on a couple of special cases. Clearly, cases with $n_s=j, n_l=q$ are always a solution for Eq.~(\ref{symmetry_condition}).
If $j,q$ are also co-prime, it means there will be only two symmetric axes, at two opposing ends of the ring. If $j$ and $q$ are not co-prime and share a non-trivial common factor $k$, then $n_s=j/k, n_l=q/k$ is a solution, and the symmetry points are spaced by $\pi/k$. These states are in the quasi-Bloch form whose period is $2\pi/k$ instead of $2\pi/q$, a $q/k$ multiple of the lattice period \cite{smith04, holland05}. For example, $q=6, j=3$ would result in a state with its density modulated with period $2\pi/3$, although we do not include a plot here. The above symmetry categories still apply to these cases.

\section{Effect of Lattice on Eigenstates} \label{Sec:Lattice_Eigenstates}

\begin{figure}[t]
\centering
\includegraphics[width=\columnwidth]{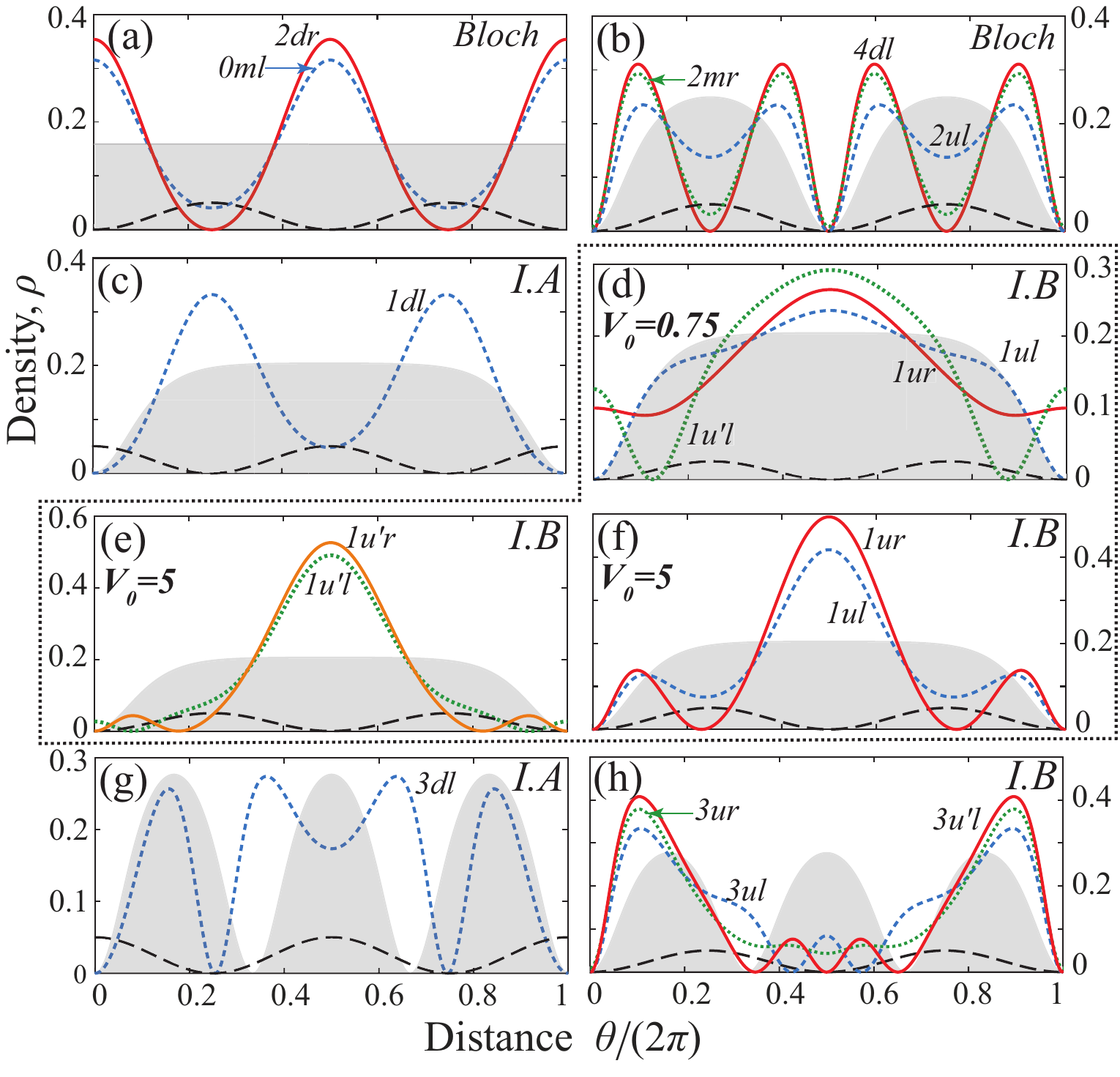}
\caption{(Color online) The nonlinear eigenstates for lattice periodicity $q=2$ in the presence of the lattice (long dashed lines). The lattice depth is $V_0=5$ and state labels correspond to those in Fig.~\ref{Fig3-nonlinear-spectrum}(f) except for panel (d) tied to Fig.~\ref{Fig3-nonlinear-spectrum}(e). Eigenstates in the absence of the lattice are shown in filled gray. (a,b) Nonlinear Bloch waves, where state and lattice periodicities match. The remaining panels are soliton solutions of various types, with the symmetry type labelled. Panels (e) and (f) represent states at a higher lattice depth $V_0=5$ on spectral branches that arise from a single branch at a lower lattice depth $V_0=0.75$, the states for which are shown in (d). }
\label{Fig5-statesM2}
\end{figure}

\begin{figure}[t]
\centering
\includegraphics[width=\columnwidth]{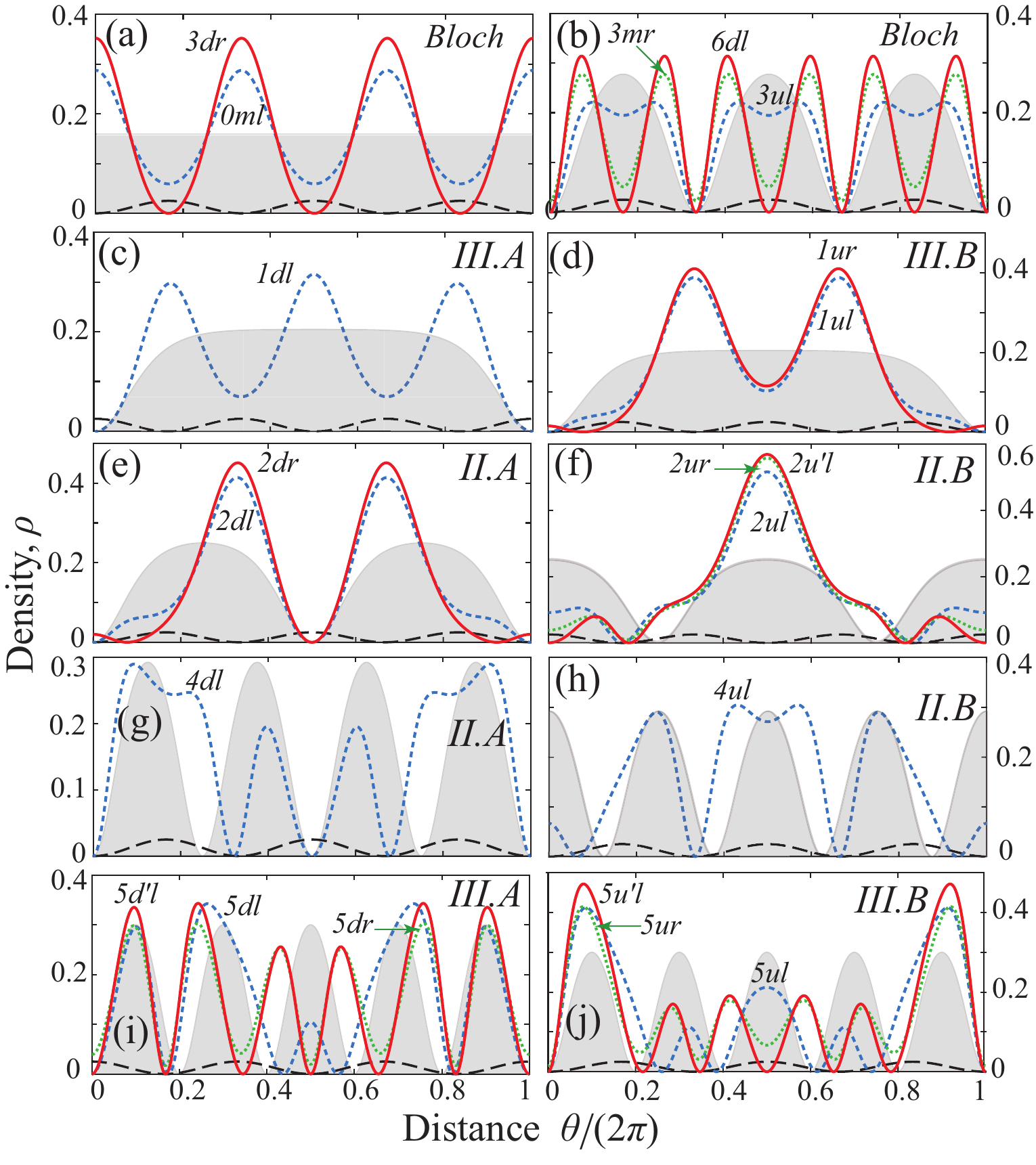}
	\caption{The nonlinear eigenstates similar to Fig.~\ref{Fig5-statesM2} but for lattice periodicity $q=3$.  The state labels correspond to Fig.~\ref{Fig3-nonlinear-spectrum}(i) with lattice depth $V_0=6$. (a,b) Nonlinear Bloch waves, where state and lattice periodicities match. For the remaining panels, the left ones belong to symmetry Type A and right ones belong to symmetry Type B.}
	\label{Fig6-statesM3}
\end{figure}

We display several states in Figs.~\ref{Fig5-statesM2} and \ref{Fig6-statesM3}, corresponding the spectra shown in Fig.~\ref{Fig3-nonlinear-spectrum}. However, we should reiterate that many of these states are accessible only by introducing rotation. For example, due to the phase boundary condition $\delta \phi/(2\pi)=n$, states $2dl$ and $2ul$ are allowed stationary states without rotation for Fig.~\ref{Fig5-statesM2} whereas, the corresponding states  in Fig.~\ref{Fig6-statesM3} are not.

The states on the main band have the Bloch structure where the periodicity of the density is commensurate with that of the lattice. On each band the states vary in shape continuously as we progress along the spectral curve, with different number of nodes at the two ends, spanned by nodeless solutions in between. This can be seen in Fig.~\ref{Fig5-statesM2}(a,b) as we progress from solution $0ml$ with no nodes to $2dr$ with 2 nodes, and from $2ul$ with 2 nodes to $4dl$ with 4 nodes, and likewise Fig.~\ref{Fig6-statesM3}(a,b) from $0ml$ with no nodes to $3dr$ with three nodes and from $3ul$ with 3 nodes to $6dl$ with 6 nodes. The progressive morphing of the solutions in between can be seen with solutions $2mr$ and $3mr$ in the respective figures.

As the lattice splits the swallowtails into two branches, at those branches with index commensurate with the lattice period, the split corresponds to a band gap and the lower branch initiates from lattice-free solution with a density minimum aligned with a lattice maximum, and the upper branch with density maximum aligned with lattice maximum. This can be see with $2dl$ and $2ul$ in Fig.~\ref{Fig5-statesM2}(a,b) and $3dl$ and $3ul$ in Fig.~\ref{Fig6-statesM3}(a,b). The grey filled shapes outline states in the absence of a lattice, however in panels (a) in both figures, we show the nodeless solution that morphs to $0ml$ in the presence of the lattice, but the $2dl$ and $3dl$ solutions can be understood to emerge from the lattice-free solution shown in panels (b) in both figures but shifted by half a period to have the density minima line up with the lattice maxima.

It is interesting to note that this pattern is maintained as we go from the state with nodes at the lower edge of the band to that at the upper edge. The number of nodes increase by $q$ by creating new minima.  We observe these in transitions from  $2ul$ to $4dl$ in Fig.~\ref{Fig5-statesM2}(b) and $3ul$ and $6dl$ in Fig.~\ref{Fig6-statesM3}(b).

In the case of intra-band solitons, we still find that the lower branch originates in a state with at least one density minimum  aligned with a lattice maximum corresponding to Type A in our symmetry classification, whereas the upper branch originates with the same lattice free state shifted and falls in with Type B. This can be seen both Figs.~\ref{Fig5-statesM2} and \ref{Fig6-statesM3}. The primary difference with the non-linear Bloch bands is that for them, the pattern repeats commensurate with the lattice period.  With deepening lattice, the incommensurate solitons originating with intraband swallowtails can get markedly more localized as seen in Figs.~\ref{Fig5-statesM2}(d,ef,g) and \ref{Fig6-statesM3}(d,e,f,j).

The localization of soliton state with incommensurate lattice can be understood by considering its Fourier modes. The Fourier expansion of soliton with nodes arising from the branch $j$ in the absence of a lattice, in Eq.~(\ref{eq2.1-1}) has the form \cite{byrd13book},
\begin{equation}
\label{eqfourier}
\psi_j(\theta)=\sqrt{\frac{2\pi}{K^2-EK}}\sum\limits_{n=0}^{\infty}\frac{q^{n+\frac{1}{2}}}{1-q^{2n+1}}\sin[j(n+\textstyle{\frac{1}{2}})\theta]
\end{equation}
where the modes included are commensurate with the branch index $j$. Introduction of the lattice couples each Fourier mode $n$ in a ladder stepped by multiples of the lattice period $n\pm mq$. If $j$ is a multiple of $q$, the coupling will fill the modes with step $q$, resulting in a Bloch form solution. If $j$ is not a multiple of $q$ but shares common factor $k$ with $q$, the modes will be coupled in steps of $k$, resulting in a quasi-Bloch state with period $2\pi/k$. If $j$ and $q$ are co-prime, the lattice coupling can spread the modes over all integers. As it spreads out in the Fourier space, there will be progressive localization in position space.

\begin{figure}[t]
\includegraphics[width=\columnwidth]{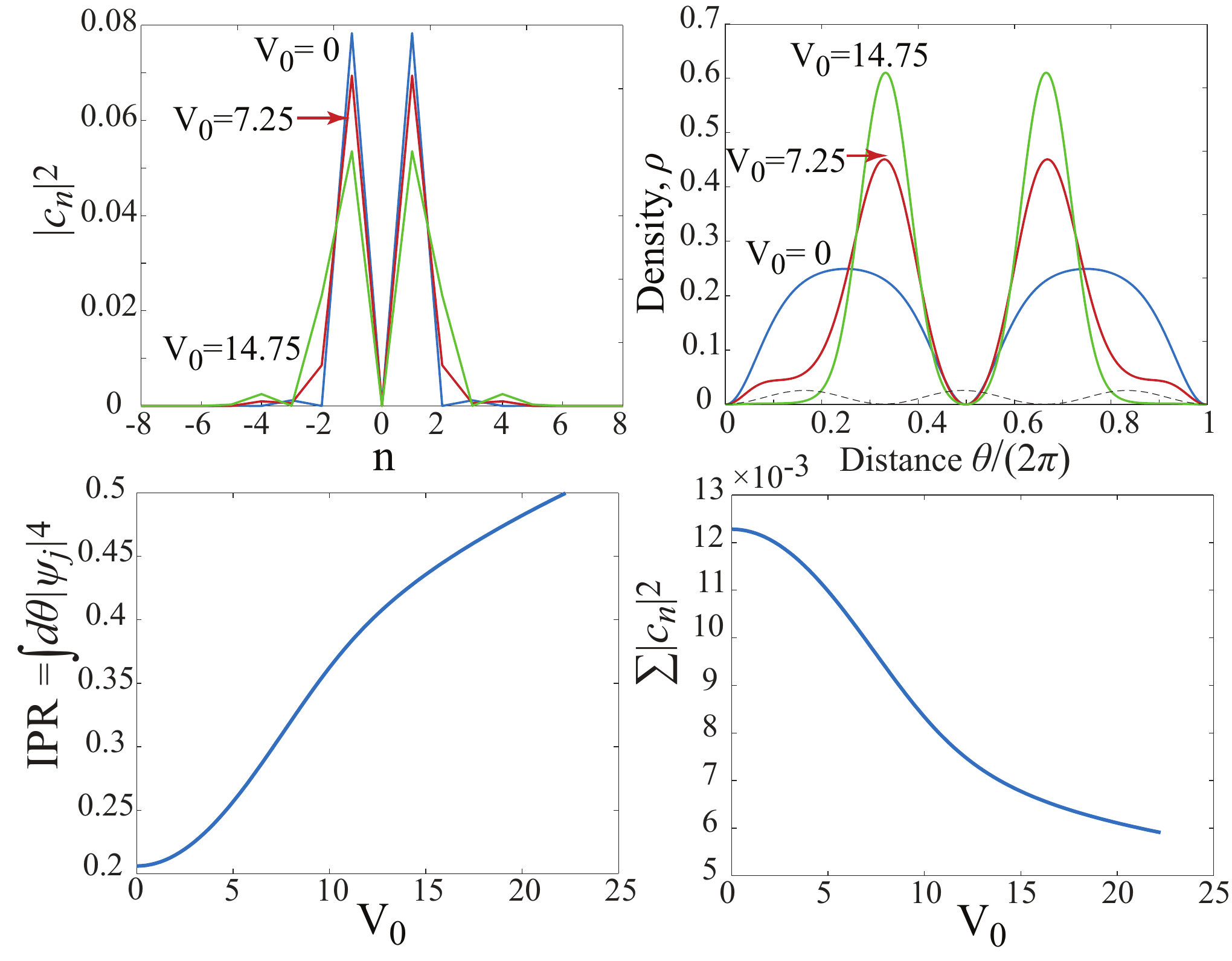}
\caption{Localization with increasing lattice depth $V_0$ is shown for lattice minima $q=3$ incommensurate with state index $j=2$. With increasing $V_0$,(a) more Fourier components have contributing amplitudes $c_n$,  (b) the mode gets more localized in position space, (c) the position space inverse participation ratio (IPR) gets larger and (d) the momentum space IPR gets smaller. In panel (b) the lattice is shown in dashed line.}
\label{Fig-localization}
\end{figure}

We illustrate these considerations in Fig.~\ref{Fig-localization}, where we present an incommensurate case with $j=2$ mode in a ring lattice with $q=3$ minima. The Fourier expansion coefficients $c_n$ are as defined in Eq.~(\ref{GPE_PW}). As the lattice depth is increased, more Fourier modes are found to be involved in panel (a), even though with lesser weights. This results in more localization in position space as shown in panel (b).  The degree of localization can be quantified by the inverse participation ratio (IPR), $\int d\theta |\psi_j|^4$ for normalized states. It is seen to climb up with deepening lattice in panel (c) indicating more localization, while its momentum space counterpart $\sum_n|c_n|^4$ in panel (d) gets progressively smaller as can be expected from the Heisenberg Uncertainty Principle.

Specifically, we can see that in Fig.~\ref{Fig-localization}(a), for the displayed mode $j=2$, in the absence of a lattice the $V_0=0$, only the modes $j(n+\frac{1}{2})=\pm 1, \pm 3, \cdots$ are present. As the lattice is turned on with $q=3$, it couples each of these modes in ladder steps of $\pm q$ as indicated in Eq.~(\ref{GPE_PW}), for example, $\pm 1\rightarrow \pm 1\pm 3=-2,3,\pm 4$ all of which acquire visible occupation in that plot as the lattice depth increases to $V_0=7.25$ and then to $V_0=14.75$.

This also underscores that for small lattice sizes on a ring and weak lattice strengths, differentiation between localized and delocalized states may not be very obvious, as can be gathered from the states shown in Figs.~\ref{Fig5-statesM2} and \ref{Fig6-statesM3}. That may only become prominent as the lattice depth is increased. However, our criterion above provides a concrete way to distinguish the states that can become localized versus those that will not.

\section{Varying Nonlinearity at Fixed lattice}

The nonlinearity and the lattice introduce complementary effects and it is interesting to compare the results of our approach where we start from the nonlinear solutions with no lattice, with those of Ref.~\cite{Guilleumas-nonlinear_ring} where the starting point was linear Bloch waves. We have confirmed that the approach taken here is completely consistent with all of the results in that study, and here we summarize a brief comparison.

\begin{figure}[t]
	\includegraphics[width=1\columnwidth]{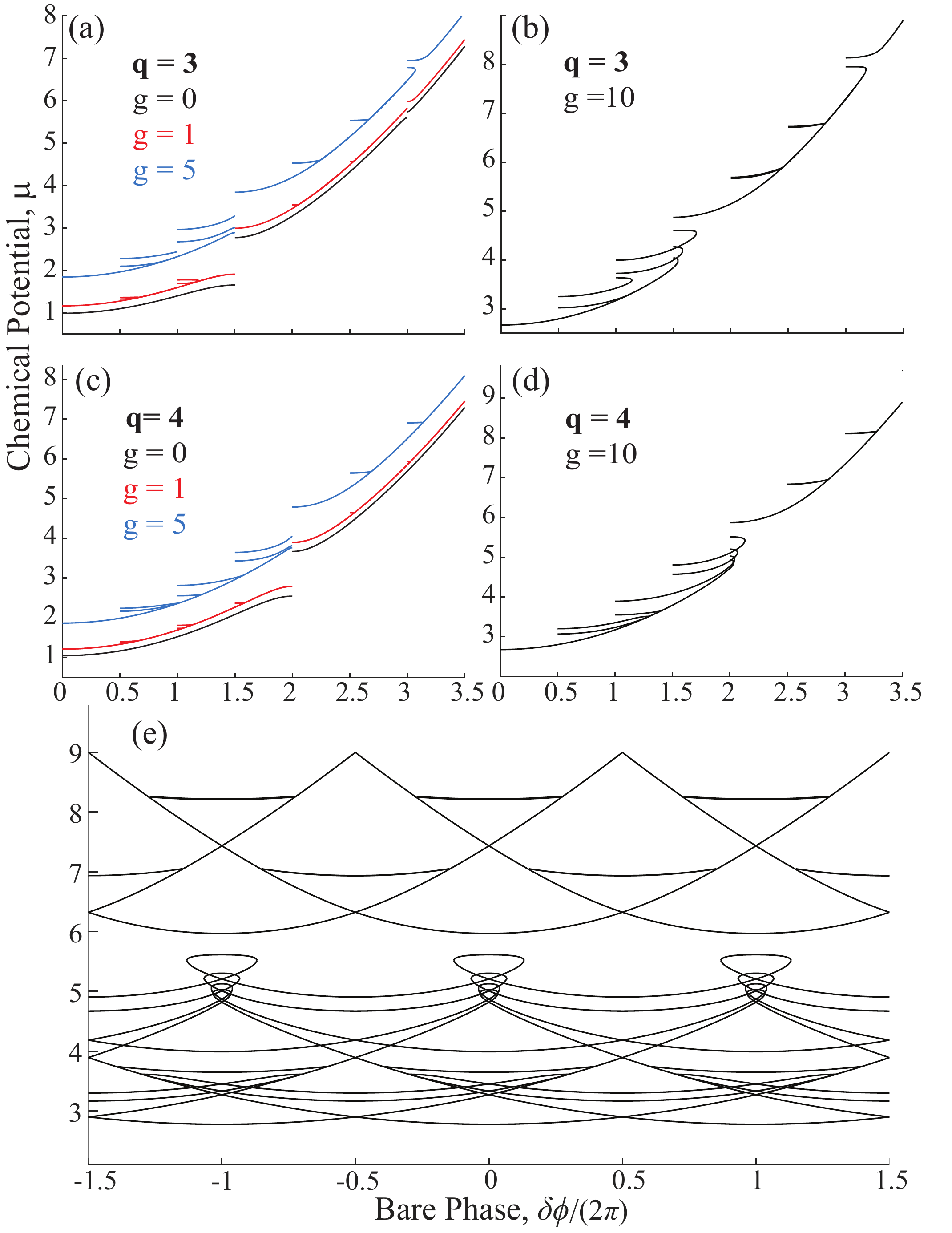}
	\caption{The chemical potential as a function of the phase is plotted for different interaction strengths for lattice periodicities $q=3$ and $q=4$, for fixed lattice strength $V_0=2.25$. In the left panels (a,c) the soliton branches appear and lengthen with increasing $g$.   At larger nonlinearity, in the right panels (b,d), the hook like structures appear that extend between the band edges.  Panel (e) is a periodic zone representation of panel (d).}
	\label{Fig7-varyg}
\end{figure}

The spectrum of the noninteracting system with lattice comprises of smooth monotonic curves with gaps at the band edges as shown in Fig.~\ref{Fig7-varyg}(a,c) for $g=0$. Increasing the lattice depth would widen the gap and flatten the bands. The eigenstates are Bloch waves, $\psi_{n,k}=e^{ik\theta}\sum\limits_l c_{n,l}e^{ilq\theta}$, with band index $n$ and quasimomentum $k$.
For fixed lattice depth, once interaction is introduced, as shown in Fig.~\ref{Fig7-varyg}(a,c), doublets of soliton branches emerge from the main branch terminating $\frac{\Delta\phi}{2\pi}=j/2$ for $j=1,2,3,\cdots$ except initially at $j=nq$ that are multiple of the lattice sites, where band gaps open up.

As the nonlinear strength is increased the soliton branches lengthen, and since they still terminate at the same points, their points of contact with the main branch slides up. More significantly at the band gaps, once the interaction strength $g$ exceeds some critical value for each gap, \cite{bronski01,Wu_Niu-landau-zener}, the primary dispersion curve extends beyond the band edge and forms the familiar hook like structure as we can see in Fig.~\ref{Fig7-varyg}(b,d) for $g=10$. This perspective has been analyzed in prior studies \cite{holland05, smith03, bronski01, wu02}. But, as we have shown in this paper, the introduction of the lattice into the nonlinear solutions offers a different and more unified perspective: The lattice splits all the swallowtail branches into two, but when $j=nq$, the split coincides with that band gap. The hook structures seen at the band edge have the same origins as with the intra-band swallowtails in the lattice-free limit.

In Fig.~\ref{Fig7-varyg}(e), we also show a periodic zone version of the spectrum that appears in the previous panel (d) of the same figure. This clearly shows the band separation.  The intraband swallow tail structures appear within the span of each band with the characteristic shape that gives them their names. The hook like structures in panel (d) associated with the band edge and the soliton branches in the band gaps, form loops in the extended zone schemes. If total energy was plotted instead of chemical potential, these rounded loops would have pointed edges instead, as illustrated in the inset in Fig.~\ref{Fig3-nonlinear-spectrum}(b), and would have the characteristic swallowtail shape.

\begin{figure}[t]
\includegraphics[width=\columnwidth]{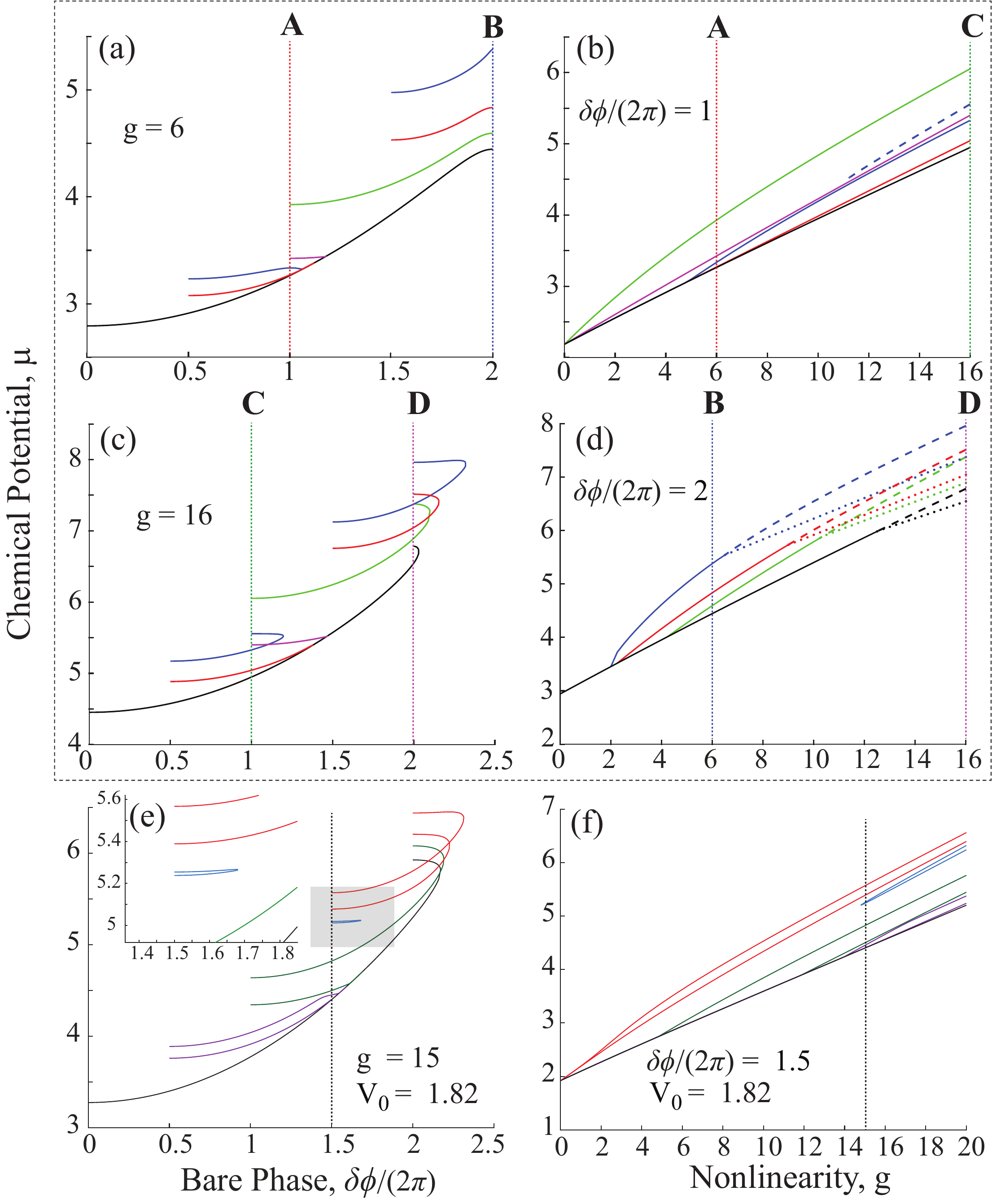}
\caption{(Color online) Comparison is made for the plots of the chemical potential as a function of the phase (a,c) at fixed interaction strength, with plots as a function of the interaction strength (b,d) at fixed phase. The vertical dotted lines mark the values in each set that are used in the other set, for example lines A and B in (a) mark the phase $\delta\phi$ values used in (b) and (d) respectively; lines A and B in (b) and (d) mark the value of $g$ used in (a). The lattice depth for (a-d) is set at $V_0=4$. The dashed lines mark intra-band soliton branches as in Fig.~3. Plots (e) and (f) are similar comparisons but specifically show the emergence of a saddlepoint bifurcation which does not originate in any of the soliton branches of the lattice-free system. }
\label{Fig8-Bifurcation}
\end{figure}

In the weak nonlinearity limit, the solitonic solutions can be viewed as superpositions of degenerate linear solutions. Under small interaction strength, such solitonic solutions will merge into nonlinear Bloch waves if $\Delta\phi$ is increased. Once the interaction strength becomes larger, these solitonic branches will extend further, until it reaches the edge of the `Brilluoin zone', where they are stopped by lattice scattering and their velocity becomes zero, and these solutions become real solutions with nodes.
For even number of lattice sites, the solitonic solutions reaching the Brilluoin zone edge will emerge as new physical solutions. These solutions are the analogs of gap solitons in literature since they emerge inside the band gaps. For odd number of lattice sites, the solutions from the solitonic branch $\delta\phi/(2\pi)=q/2-0.5$ are physical, and they exist with arbitrary small interaction strength. If the interaction strength is further increased, the nonlinear Bloch waves and the solitonic branches will extend beyond the edge of the Brilluoin zone, reaches a phase maximum, then turns back and stops at the Brilluoin zone, forming the hook structure. The solution at the lower parts of these hooks at the Brilluoin zone edge then emerge as new physical solutions with nonzero velocity.

Figure~\ref{Fig8-Bifurcation} compares our representation of the chemical potential as a function of the bare phase change around the ring in the left panels, with the variation of the chemical potential with increasing nonlinearity in the right panels.  Note that the spectrum splits as the nonlinearity increases in a qualitatively similar manner to when the lattice depth is increased in Fig.~\ref{Fig2-VaryV0}.  The vertical lines labelled in bold upper case letters mark counterparts to be matched in the two representations. In the $\mu$ versus $\delta\phi$ representation, we observe the hook structures, at line $C$ and $D$ in panel (c) and (d) respectively; this manifests itself as splitting of the branch in $\mu$ versus $g$ representation in panels (b) and (d) respectively.

The comparison shows that the splittings observed as nonlinaerity increases in Fig.~\ref{Fig2-VaryV0}(b,d,f), are of three different types. The split branches marked by solid lines mark the emergence of the equivalent of intra-band swallowtail branches.  The splits marked by dashed and dotted lines correspond to the upper and lower sections of the hook structures. Curiously, the natural continuation of each such branch that splits as $g$ increases actually corresponds to the upper end of the hook in our representation, even though in the left panels it is the lower section that is the immediate continuation of the main branch. The most unusual feature appears in Fig.~\ref{Fig2-VaryV0}(e,f) where a pair of branches appear disconnected from any of the main branches, in both representations. This is a so called saddle node bifuration that appears when a pair of fixed points emerge seemingly out of `blue sky' as such bifurcations are also sometimes referred to \cite{Guilleumas-nonlinear_ring}.

\section{Stability of States}

As noted in Sec.~\ref{Sec:Lattice_Eigenstates}, the lattice breaks the translational symmetry upto its period the splitting of the spectral branches are associated with the different relative position with respect to the lattice. Here, we show that they are associated with different dynamical stability properties. We consider small perturbation around the mean field stationary states:
\bn \psi(\theta,t)=\psi_0(\theta)+\delta u e^{-i\mu t} e^{-i\omega t}+\delta v^* e^{-i\mu t} e^{i\omega^* t}\en
and the solve the Bogoliubov equations \cite{RMP-Sringari-1999} for the normal modes of the fluctuations.
\bn (H_0+2g|\psi_0|^2-\mu)\delta u+ g\psi_0^2\delta v=w \delta u\n\\
-(H_0+2g|\psi_0|^2-\mu)\delta v+ g(\psi^*_0)^2\delta u=w \delta u\en
If the angular frequencies $\omega$ of the normal modes have  imaginary components and if ${\rm Im}(\omega)>0$ then the fluctuations would grow exponentially indicating dynamical instability.

\begin{figure}[t]
\includegraphics[width=\columnwidth]{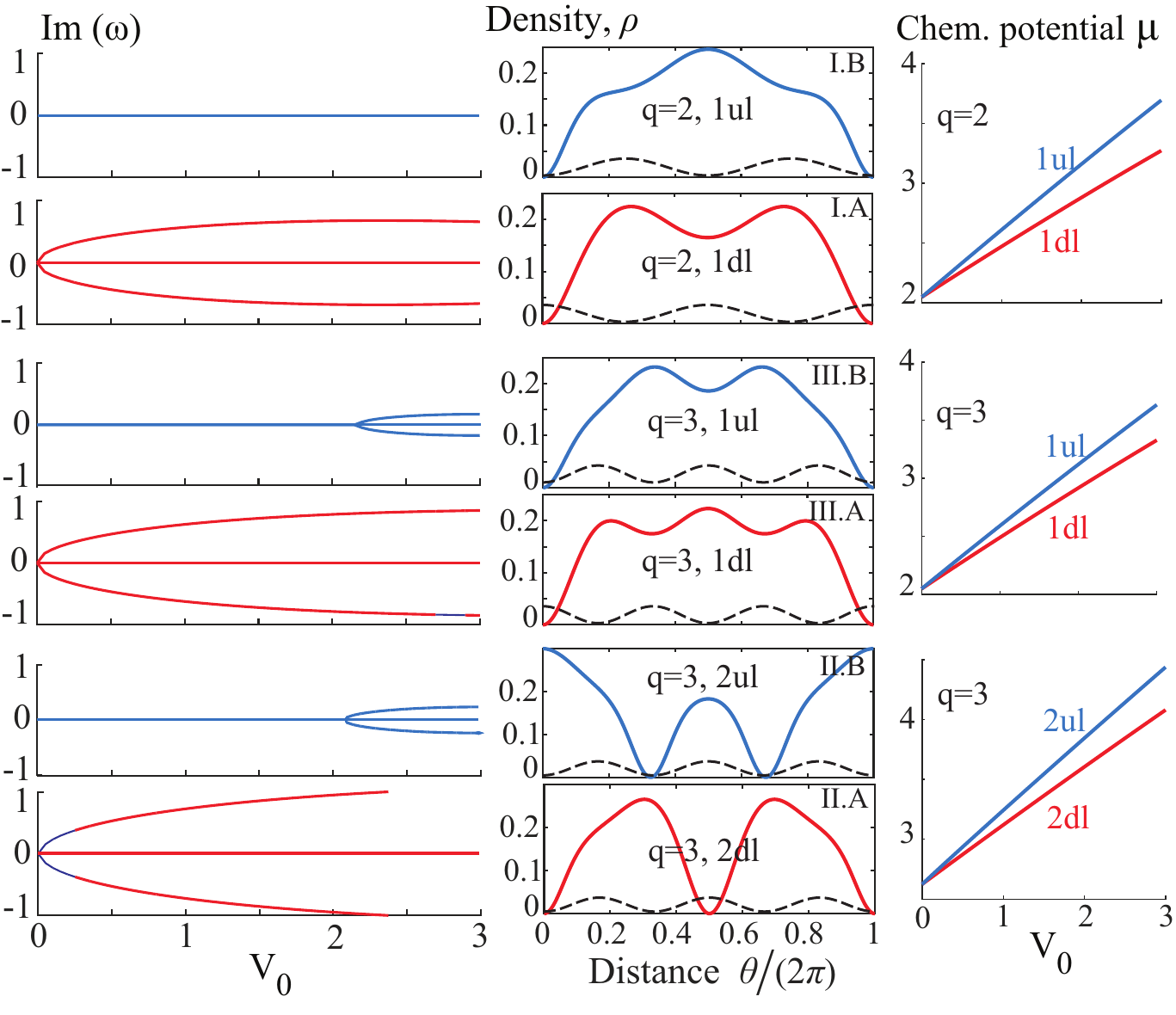}
	\caption{Examples of Bogoliubov analysis of the splitting soliton solutions. Left panels show the imaginary part of the Bogoliubov excitations as the lattice is turned on. Middle panels show the density profile of the split solutions for fixed $V_0=1$; dashed lines indicate the lattice potential. Right panels show the corresponding evolution of the chemical potential  as the lattice is turned on. }
\label{Fig9_stability}
\end{figure}

It was noted that in the absence of a lattice with repulsive interaction, both plane waves and soliton train solutions are dynamically stable \cite{carr09}. Here we studied the dynamical stabilities of the lowest soliton branches in the presence of lattice for different number lattice sites on the ring, $q=1, 2$ and  $3$. For each swallowtail branch $j$, the two split branches that emerges with the introduction of the lattice not only differ in symmetry properties, but also in the stability properties. In Fig.~\ref{Fig9_stability}, we plot the imaginary components for each branch as a function of increasing lattice depth. Rather surprisingly, for the parameter range we have tested, for each split pair, the branch with the lower chemical potential labelled as $1dl, 2dl$ are seen to be dynamically unstable as soon as the lattice is turned on indicated by the presence of imaginary components of $\omega$ even as $V_0\rightarrow 0$. In contrast such instability is manifest only at larger lattice depth for the branch with the higher chemical potential. We found that the pattern appears to hold for higher lattice sites and branches as well but the difference is not as stark and not conclusive. Simple energy considerations makes this behavior rather non-intuitive, but we do not yet have a satisfactory explanation for this behavior.

\section{Large Lattice limit}

The ring topology is particularly significant when the size is small and the number of lattice sites is few in number. As the number of sites is increased without changing the lattice period and the number of particles per site, the ring gets larger and we can consider the infinite lattice limit. This however requires a different scaling than we have used in the rest of the paper, where we have scaled length by $R$ and energy by $\hbar^2/(mR^2)$, so that keeping $V$ and $g$ constant has the effect of diminishing  the potential and the nonlinear term relative to the kinetic energy by a factor $\propto R^2$.  This did not impact our observations where comparisons were made for fixed lattice sites.  Now, to compare the variation with changing number of lattice sites, while keeping the lattice period, depth and the particle density constant, we scale the potential and the nonlinear strength by the square of the number of the lattice sites $q^2$ in Eq.~(\ref{eq1.1-9}) then divide the resulting chemical potential by $q^2$.

\begin{figure}[t]
\includegraphics[width=\columnwidth]{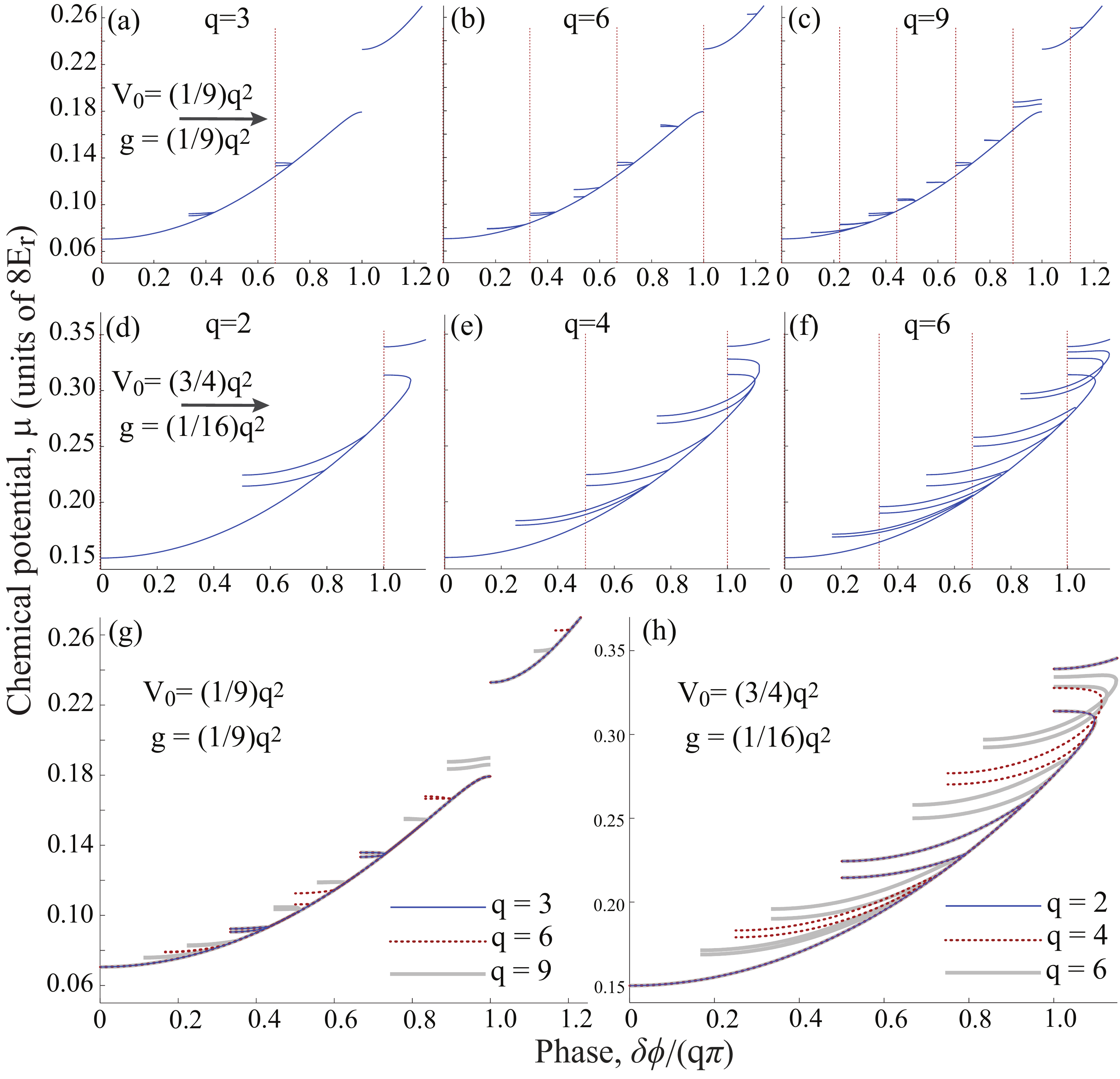}
	\caption{The effect of increasing the number $q$ of lattice sites is shown for two different values of the lattice strength $V_0$ and nonlinear constant $g$ in (a-c) and (d-f). Allowed solutions in the absence of rotation are marked by vertical dotted lines. For consistent comparison across different lattice sites, the $V_0$ and $g$ are scaled by $q^2$, and the chemical potential $\mu$ and the bare phase $\delta\phi$ are also in recoil units fixed by the lattice constant. Overlaying the ones from (a-c) in panel (g) and those from (d-f) in (h) shows that the primary dispersion curve corresponding to nonlinear Bloch waves are identical for any number of lattice sites, while soliton branches at specific multiples overlap as new ones emerge with increasing sites.}
\label{Fig10-vary-q}
\end{figure}

Some of our simulations are shown in Fig.~\ref{Fig10-vary-q} where we plot the chemical potential versus the bare phase around the ring $\delta\phi$ as before, but now scaled by a multiple of the recoil energy $8E_r=4 \pi^2\hbar^2/(md^2)$ and $q\pi$ respectively, where $d=2\pi R/q$ is the lattice constant. The phase scaling is similar to scaling the linear wave vectors by the recoil momentum $k_r=\pi/d$.  We plot two cases, one set in Fig.~\ref{Fig10-vary-q}(a-c) for $q=3,6,9$ for $V_0=q^2/9$ and $g=q^2/9$ in the units of lattice constant and recoil energy mentioned above, and another set in Fig.~\ref{Fig10-vary-q}(d-f)  for $q=2,4,6$ for $V_0=3q^2/4$ and $g=q^2/16$.  Scaled this way, we find that the shape of primary dispersion curve is identical for any number of lattice sites as long as the lattice depth $V_0$ and nonlinearity including particle density remain the same. This can be seen clearly when we overlay the plots from (a-c) in panel (g) and the plots from (d-f) in panel (h), the main bands are identical and lie on top of each other. This behavior is identical to the case without nonlinearity, and specifically we note from panels (d-f) and (g) that even the distinctive nonlienar feature, the hook structure, is identical for different number of lattice sites.

On the other hand, new soliton branches emerge with increasing lattice sites. For weak lattices, the gaps open at multiples of the lattice site number $j=nq$ and with $q-1$ pairs of intraband swallowtails for each band. When $q$ values have a common factor, they share common soliton branches and the shape of those branches is also identical for the $q$ values. This can be seen in panels Fig.~\ref{Fig10-vary-q}(a-c), for example with the two pairs of intra-band soliton branches for $q=3$, which also appear identically for $q=6$ and $q=9$,  and likewise in Fig.~\ref{Fig10-vary-q}(d-f), the single pair of intra-band soliton branches for $q=2$ also appear identically for $q=4$ and $q=6$.  This similarity is also reflected in the quantum states, with the basic shape of the wavefunction for the fundamental period $q=3$, simply repeating twice for $q=6$ over the lattice and thrice for $q=9$.

We can expect that as the number of sites in the ring gets larger, keeping the lattice depth and nonlinear strength and particle density fixed, the main dispersion curve corresponding to the nonlinear Bloch states will remain unchanged, but more and more soliton branches will appear.  However, with increasing number of such branches, the soliton branches that slide into the gap as seen in Fig.~\ref{Fig10-vary-q}(c) and (g) become more exceptional leading up to the well-studied properties of gap solitons.

\section{Conclusions and Outlook}

In a finite ring-shaped lattice, we find that the nonlinear solutions without the lattice define the landscape of solutions with the lattice. Many of the standard labels and classifications, such as referring to the solutions as solitons, are a relic of studies in the context of open lattices in the infinite and tight binding limit, and are not a particularly useful way of understanding finite closed lattices that can be possibly weak. In a ring without a lattice, there are three types of solutions: Plane waves, solutions with nodes, modulated solutions without nodes. The presence of a lattice modifies these to the following set of independent classifications of solutions: (I)  Solutions with period commensurate/incommensurate with the lattice period, and (II) solutions with/without nodes. The first classification can be sub-grouped into nonlinear Bloch waves that exactly match the periodicity of the lattice, and their super-periodic counterparts that have periods that are multiples of the lattice period.

The commensurate solutions emerge from the plane wave solutions and the swallowtails at multiples of the lattice period.  The incommensurate ones originate from intra-band swallowtails. With increasing lattice strength, the commensurate ones will remain delocalized, whereas with increasing lattice depth, the incommensurate ones will tend to localize.

As already noted in Ref.~\cite{Guilleumas-nonlinear_ring}, the concept of gap solitons is not very distinctive in a finite ring lattice.  There are solutions that exist in the band gap that originate with the migration of the intra-band swallowtails with increasing lattice depth, but their behavior falls into the general classification above.

The most significant aspect of the solutions of a ring lattice is the ability to access a full range of solutions of the infinite lattice by using rotation.  This makes the ring lattice an invaluable and unique tool for the study of nonlinear dynamics. The continuous transition between solutions possible with rotation can be utilized to probe and modify the dynamics of the system, to access vastly different kinds of behaviour within the same system.  We have already observed that even in the linear case \cite{Das-Brooks-Brattley}, and in our continuing work we anticipate even richer dynamics in the nonlinear regime. An exploration of the negative nonlinearity constitutes a natural continuation of this work. Finally, there is much potential for exploring the quantum dynamics in ring lattices beyond the mean field regime, which also forms part of our ongoing research.

\begin{acknowledgments} We gratefully acknowledge valuable discussions with D. Schneble, and the support of the NSF under Grant No. PHY-2011767. \end{acknowledgments}

\vfill


\begin{thebibliography}{51}
\expandafter\ifx\csname natexlab\endcsname\relax\def\natexlab#1{#1}\fi
\expandafter\ifx\csname bibnamefont\endcsname\relax
  \def\bibnamefont#1{#1}\fi
\expandafter\ifx\csname bibfnamefont\endcsname\relax
  \def\bibfnamefont#1{#1}\fi
\expandafter\ifx\csname citenamefont\endcsname\relax
  \def\citenamefont#1{#1}\fi
\expandafter\ifx\csname url\endcsname\relax
  \def\url#1{\texttt{#1}}\fi
\expandafter\ifx\csname urlprefix\endcsname\relax\def\urlprefix{URL }\fi
\providecommand{\bibinfo}[2]{#2}
\providecommand{\eprint}[2][]{\url{#2}}

\bibitem[{\citenamefont{Bloch et~al.}(2008)\citenamefont{Bloch, Dalibard, and
  Zwerger}}]{Bloch-RMP-Many-Body}
\bibinfo{author}{\bibfnamefont{I.}~\bibnamefont{Bloch}},
  \bibinfo{author}{\bibfnamefont{J.}~\bibnamefont{Dalibard}}, \bibnamefont{and}
  \bibinfo{author}{\bibfnamefont{W.}~\bibnamefont{Zwerger}},
  \bibinfo{journal}{Rev. Mod. Phys.} \textbf{\bibinfo{volume}{80}},
  \bibinfo{pages}{885} (\bibinfo{year}{2008}).

\bibitem[{\citenamefont{Morsch and Oberthaler}(2006)}]{RMP-Oberthaler}
\bibinfo{author}{\bibfnamefont{O.}~\bibnamefont{Morsch}} \bibnamefont{and}
  \bibinfo{author}{\bibfnamefont{M.}~\bibnamefont{Oberthaler}},
  \bibinfo{journal}{Rev. Mod. Phys.} \textbf{\bibinfo{volume}{78}},
  \bibinfo{pages}{179} (\bibinfo{year}{2006}).

\bibitem[{\citenamefont{Greiner et~al.}(2002)\citenamefont{Greiner, Mandel,
  Esslinger, Hansch, and Bloch}}]{Bloch-mott}
\bibinfo{author}{\bibfnamefont{M.}~\bibnamefont{Greiner}},
  \bibinfo{author}{\bibfnamefont{O.}~\bibnamefont{Mandel}},
  \bibinfo{author}{\bibfnamefont{T.}~\bibnamefont{Esslinger}},
  \bibinfo{author}{\bibfnamefont{T.~W.} \bibnamefont{Hansch}},
  \bibnamefont{and} \bibinfo{author}{\bibfnamefont{I.}~\bibnamefont{Bloch}},
  \bibinfo{journal}{Nature} \textbf{\bibinfo{volume}{415}}, \bibinfo{pages}{39}
  (\bibinfo{year}{2002}).

\bibitem[{\citenamefont{Dalfovo et~al.}(1999)\citenamefont{Dalfovo, Giorgini,
  Pitaevskii, and Stringari}}]{RMP-Sringari-1999}
\bibinfo{author}{\bibfnamefont{F.}~\bibnamefont{Dalfovo}},
  \bibinfo{author}{\bibfnamefont{S.}~\bibnamefont{Giorgini}},
  \bibinfo{author}{\bibfnamefont{L.~P.} \bibnamefont{Pitaevskii}},
  \bibnamefont{and}
  \bibinfo{author}{\bibfnamefont{S.}~\bibnamefont{Stringari}},
  \bibinfo{journal}{Rev. Mod. Phys.} \textbf{\bibinfo{volume}{71}},
  \bibinfo{pages}{463} (\bibinfo{year}{1999}).

\bibitem[{\citenamefont{Ramanathan et~al.}(2011)\citenamefont{Ramanathan,
  Wright, Muniz, Zelan, Hill, Lobb, Helmerson, Phillips, and
  Campbell}}]{ramanathan}
\bibinfo{author}{\bibfnamefont{A.}~\bibnamefont{Ramanathan}},
  \bibinfo{author}{\bibfnamefont{K.~C.} \bibnamefont{Wright}},
  \bibinfo{author}{\bibfnamefont{S.~R.} \bibnamefont{Muniz}},
  \bibinfo{author}{\bibfnamefont{M.}~\bibnamefont{Zelan}},
  \bibinfo{author}{\bibfnamefont{W.~T.} \bibnamefont{Hill}},
  \bibinfo{author}{\bibfnamefont{C.~J.} \bibnamefont{Lobb}},
  \bibinfo{author}{\bibfnamefont{K.}~\bibnamefont{Helmerson}},
  \bibinfo{author}{\bibfnamefont{W.~D.} \bibnamefont{Phillips}},
  \bibnamefont{and} \bibinfo{author}{\bibfnamefont{G.~K.}
  \bibnamefont{Campbell}}, \bibinfo{journal}{Phys. Rev. Lett.}
  \textbf{\bibinfo{volume}{106}}, \bibinfo{pages}{130401}
  (\bibinfo{year}{2011}).

\bibitem[{\citenamefont{Wright et~al.}(2013)\citenamefont{Wright, Blakestad,
  Lobb, Phillips, and Campbell}}]{Phillips_Campbell_superfluid_2013}
\bibinfo{author}{\bibfnamefont{K.~C.} \bibnamefont{Wright}},
  \bibinfo{author}{\bibfnamefont{R.~B.} \bibnamefont{Blakestad}},
  \bibinfo{author}{\bibfnamefont{C.~J.} \bibnamefont{Lobb}},
  \bibinfo{author}{\bibfnamefont{W.~D.} \bibnamefont{Phillips}},
  \bibnamefont{and} \bibinfo{author}{\bibfnamefont{G.~K.}
  \bibnamefont{Campbell}}, \bibinfo{journal}{Phys. Rev. Lett.}
  \textbf{\bibinfo{volume}{110}}, \bibinfo{pages}{025302}
  (\bibinfo{year}{2013}).

\bibitem[{\citenamefont{Franke-Arnold et~al.}(2007)\citenamefont{Franke-Arnold,
  Leach, Padgett, Lembessis, Ellinas, Wright, Girkin, \"{O}hberg, and
  Arnold}}]{Padgett}
\bibinfo{author}{\bibfnamefont{S.}~\bibnamefont{Franke-Arnold}},
  \bibinfo{author}{\bibfnamefont{J.}~\bibnamefont{Leach}},
  \bibinfo{author}{\bibfnamefont{M.~J.} \bibnamefont{Padgett}},
  \bibinfo{author}{\bibfnamefont{V.~E.} \bibnamefont{Lembessis}},
  \bibinfo{author}{\bibfnamefont{D.}~\bibnamefont{Ellinas}},
  \bibinfo{author}{\bibfnamefont{A.~J.} \bibnamefont{Wright}},
  \bibinfo{author}{\bibfnamefont{J.~M.} \bibnamefont{Girkin}},
  \bibinfo{author}{\bibfnamefont{P.}~\bibnamefont{\"{O}hberg}},
  \bibnamefont{and} \bibinfo{author}{\bibfnamefont{A.~S.}
  \bibnamefont{Arnold}}, \bibinfo{journal}{Opt. Express}
  \textbf{\bibinfo{volume}{15}}, \bibinfo{pages}{8619} (\bibinfo{year}{2007}).

\bibitem[{\citenamefont{Zambrini and Barnett}(2007)}]{Zambrini:07}
\bibinfo{author}{\bibfnamefont{R.}~\bibnamefont{Zambrini}} \bibnamefont{and}
  \bibinfo{author}{\bibfnamefont{S.~M.} \bibnamefont{Barnett}},
  \bibinfo{journal}{Opt. Express} \textbf{\bibinfo{volume}{15}},
  \bibinfo{pages}{15214} (\bibinfo{year}{2007}).

\bibitem[{\citenamefont{Das}(2020)}]{Das-PRL-localization}
\bibinfo{author}{\bibfnamefont{K.~K.} \bibnamefont{Das}},
  \bibinfo{journal}{Phys. Rev. Lett.} \textbf{\bibinfo{volume}{125}},
  \bibinfo{pages}{070401} (\bibinfo{year}{2020}).

\bibitem[{\citenamefont{Kol\'a\ifmmode~\check{r}\else \v{r}\fi{}
  et~al.}(2015)\citenamefont{Kol\'a\ifmmode~\check{r}\else \v{r}\fi{},
  Opatrn\'y, and Das}}]{Opatrny-Kolar-Das-rotation}
\bibinfo{author}{\bibfnamefont{M.}~\bibnamefont{Kol\'a\ifmmode~\check{r}\else
  \v{r}\fi{}}}, \bibinfo{author}{\bibfnamefont{T.}~\bibnamefont{Opatrn\'y}},
  \bibnamefont{and} \bibinfo{author}{\bibfnamefont{K.~K.} \bibnamefont{Das}},
  \bibinfo{journal}{Phys. Rev. A} \textbf{\bibinfo{volume}{92}},
  \bibinfo{pages}{043630} (\bibinfo{year}{2015}).

\bibitem[{\citenamefont{Brooks et~al.}(2021)\citenamefont{Brooks, Brattley, and
  Das}}]{Das-Brooks-Brattley}
\bibinfo{author}{\bibfnamefont{C.}~\bibnamefont{Brooks}},
  \bibinfo{author}{\bibfnamefont{A.}~\bibnamefont{Brattley}}, \bibnamefont{and}
  \bibinfo{author}{\bibfnamefont{K.~K.} \bibnamefont{Das}},
  \bibinfo{journal}{Phys. Rev. A} \textbf{\bibinfo{volume}{103}},
  \bibinfo{pages}{013322} (\bibinfo{year}{2021}).

\bibitem[{\citenamefont{Aghamalyan et~al.}(2013)\citenamefont{Aghamalyan,
  Amico, and Kwek}}]{Aghamalyan-two-ring-lattice}
\bibinfo{author}{\bibfnamefont{D.}~\bibnamefont{Aghamalyan}},
  \bibinfo{author}{\bibfnamefont{L.}~\bibnamefont{Amico}}, \bibnamefont{and}
  \bibinfo{author}{\bibfnamefont{L.~C.} \bibnamefont{Kwek}},
  \bibinfo{journal}{Phys. Rev. A} \textbf{\bibinfo{volume}{88}},
  \bibinfo{pages}{063627} (\bibinfo{year}{2013}).

\bibitem[{\citenamefont{Satija et~al.}(2013)\citenamefont{Satija, Pando~L., and
  Tiesinga}}]{Tiesinga-soliton-lattice}
\bibinfo{author}{\bibfnamefont{I.~I.} \bibnamefont{Satija}},
  \bibinfo{author}{\bibfnamefont{C.~L.} \bibnamefont{Pando~L.}},
  \bibnamefont{and} \bibinfo{author}{\bibfnamefont{E.}~\bibnamefont{Tiesinga}},
  \bibinfo{journal}{Phys. Rev. A} \textbf{\bibinfo{volume}{87}},
  \bibinfo{pages}{033608} (\bibinfo{year}{2013}).

\bibitem[{\citenamefont{Jezek and Cataldo}(2011)}]{Jezek-winding-number}
\bibinfo{author}{\bibfnamefont{D.~M.} \bibnamefont{Jezek}} \bibnamefont{and}
  \bibinfo{author}{\bibfnamefont{H.~M.} \bibnamefont{Cataldo}},
  \bibinfo{journal}{Phys. Rev. A} \textbf{\bibinfo{volume}{83}},
  \bibinfo{pages}{013629} (\bibinfo{year}{2011}).

\bibitem[{\citenamefont{Nigro et~al.}(2018)\citenamefont{Nigro, Capuzzi, and
  Jezek}}]{Nigro_2018}
\bibinfo{author}{\bibfnamefont{M.}~\bibnamefont{Nigro}},
  \bibinfo{author}{\bibfnamefont{P.}~\bibnamefont{Capuzzi}}, \bibnamefont{and}
  \bibinfo{author}{\bibfnamefont{D.~M.} \bibnamefont{Jezek}},
  \bibinfo{journal}{Phys. Rev. A} \textbf{\bibinfo{volume}{98}},
  \bibinfo{pages}{063622} (\bibinfo{year}{2018}).

\bibitem[{\citenamefont{Opatrn\'y et~al.}(2015)\citenamefont{Opatrn\'y,
  Kol\'a\v{r}, and Das}}]{Opatrny-Kolar-Das-LMG}
\bibinfo{author}{\bibfnamefont{T.}~\bibnamefont{Opatrn\'y}},
  \bibinfo{author}{\bibfnamefont{M.}~\bibnamefont{Kol\'a\v{r}}},
  \bibnamefont{and} \bibinfo{author}{\bibfnamefont{K.~K.} \bibnamefont{Das}},
  \bibinfo{journal}{Phys. Rev. A} \textbf{\bibinfo{volume}{91}},
  \bibinfo{pages}{053612} (\bibinfo{year}{2015}).

\bibitem[{\citenamefont{Das and Christ}(2019)}]{Das-Christ}
\bibinfo{author}{\bibfnamefont{K.~K.} \bibnamefont{Das}} \bibnamefont{and}
  \bibinfo{author}{\bibfnamefont{J.}~\bibnamefont{Christ}},
  \bibinfo{journal}{Phys. Rev. A} \textbf{\bibinfo{volume}{99}},
  \bibinfo{pages}{013604} (\bibinfo{year}{2019}).

\bibitem[{\citenamefont{Arwas and Cohen}(2016)}]{Doron-Cohen-1}
\bibinfo{author}{\bibfnamefont{G.}~\bibnamefont{Arwas}} \bibnamefont{and}
  \bibinfo{author}{\bibfnamefont{D.}~\bibnamefont{Cohen}},
  \bibinfo{journal}{New Journal of Physics} \textbf{\bibinfo{volume}{18}},
  \bibinfo{pages}{015007} (\bibinfo{year}{2016}).

\bibitem[{\citenamefont{Hettiarachchilage
  et~al.}(2013)\citenamefont{Hettiarachchilage, Rousseau, Tam, Jarrell, and
  Moreno}}]{Moreno-Bose-Hubbard}
\bibinfo{author}{\bibfnamefont{K.}~\bibnamefont{Hettiarachchilage}},
  \bibinfo{author}{\bibfnamefont{V.~G.} \bibnamefont{Rousseau}},
  \bibinfo{author}{\bibfnamefont{K.-M.} \bibnamefont{Tam}},
  \bibinfo{author}{\bibfnamefont{M.}~\bibnamefont{Jarrell}}, \bibnamefont{and}
  \bibinfo{author}{\bibfnamefont{J.}~\bibnamefont{Moreno}},
  \bibinfo{journal}{Phys. Rev. A} \textbf{\bibinfo{volume}{87}},
  \bibinfo{pages}{051607} (\bibinfo{year}{2013}).

\bibitem[{\citenamefont{Cataldo and
  Jezek}(2011)}]{Jezek-Bose-Hubbard-ring-lattice}
\bibinfo{author}{\bibfnamefont{H.~M.} \bibnamefont{Cataldo}} \bibnamefont{and}
  \bibinfo{author}{\bibfnamefont{D.~M.} \bibnamefont{Jezek}},
  \bibinfo{journal}{Phys. Rev. A} \textbf{\bibinfo{volume}{84}},
  \bibinfo{pages}{013602} (\bibinfo{year}{2011}).

\bibitem[{\citenamefont{Maik et~al.}(2011)\citenamefont{Maik, Buonsante,
  Vezzani, and Zakrzewski}}]{Maik-dipolar}
\bibinfo{author}{\bibfnamefont{M.}~\bibnamefont{Maik}},
  \bibinfo{author}{\bibfnamefont{P.}~\bibnamefont{Buonsante}},
  \bibinfo{author}{\bibfnamefont{A.}~\bibnamefont{Vezzani}}, \bibnamefont{and}
  \bibinfo{author}{\bibfnamefont{J.}~\bibnamefont{Zakrzewski}},
  \bibinfo{journal}{Phys. Rev. A} \textbf{\bibinfo{volume}{84}},
  \bibinfo{pages}{053615} (\bibinfo{year}{2011}).

\bibitem[{\citenamefont{Pinheiro and de~Toledo~Piza}(2013)}]{Piza-ring-lattice}
\bibinfo{author}{\bibfnamefont{F.}~\bibnamefont{Pinheiro}} \bibnamefont{and}
  \bibinfo{author}{\bibfnamefont{A.~F.~R.} \bibnamefont{de~Toledo~Piza}},
  \bibinfo{journal}{Journal of Physics B: Atomic, Molecular and Optical
  Physics} \textbf{\bibinfo{volume}{46}}, \bibinfo{pages}{205303}
  (\bibinfo{year}{2013}).

\bibitem[{\citenamefont{Polo et~al.}(2020)\citenamefont{Polo, Naldesi,
  Minguzzi, and Amico}}]{Minguzzi-PRA-two-bosons}
\bibinfo{author}{\bibfnamefont{J.}~\bibnamefont{Polo}},
  \bibinfo{author}{\bibfnamefont{P.}~\bibnamefont{Naldesi}},
  \bibinfo{author}{\bibfnamefont{A.}~\bibnamefont{Minguzzi}}, \bibnamefont{and}
  \bibinfo{author}{\bibfnamefont{L.}~\bibnamefont{Amico}},
  \bibinfo{journal}{Phys. Rev. A} \textbf{\bibinfo{volume}{101}},
  \bibinfo{pages}{043418} (\bibinfo{year}{2020}).

\bibitem[{\citenamefont{Richaud and Penna}(2019)}]{Penna}
\bibinfo{author}{\bibfnamefont{A.}~\bibnamefont{Richaud}} \bibnamefont{and}
  \bibinfo{author}{\bibfnamefont{V.}~\bibnamefont{Penna}},
  \bibinfo{journal}{Phys. Rev. A} \textbf{\bibinfo{volume}{100}},
  \bibinfo{pages}{013609} (\bibinfo{year}{2019}).

\bibitem[{\citenamefont{Aghamalyan et~al.}(2015)\citenamefont{Aghamalyan,
  Cominotti, Rizzi, Rossini, Hekking, Minguzzi, Kwek, and
  Amico}}]{Aghamalyan-AQUID}
\bibinfo{author}{\bibfnamefont{D.}~\bibnamefont{Aghamalyan}},
  \bibinfo{author}{\bibfnamefont{M.}~\bibnamefont{Cominotti}},
  \bibinfo{author}{\bibfnamefont{M.}~\bibnamefont{Rizzi}},
  \bibinfo{author}{\bibfnamefont{D.}~\bibnamefont{Rossini}},
  \bibinfo{author}{\bibfnamefont{F.}~\bibnamefont{Hekking}},
  \bibinfo{author}{\bibfnamefont{A.}~\bibnamefont{Minguzzi}},
  \bibinfo{author}{\bibfnamefont{L.-C.} \bibnamefont{Kwek}}, \bibnamefont{and}
  \bibinfo{author}{\bibfnamefont{L.}~\bibnamefont{Amico}},
  \bibinfo{journal}{New Journal of Physics} \textbf{\bibinfo{volume}{17}},
  \bibinfo{pages}{045023} (\bibinfo{year}{2015}).

\bibitem[{\citenamefont{Arwas et~al.}(2017)\citenamefont{Arwas, Cohen, Hekking,
  and Minguzzi}}]{Minguzzi-resonant-persistent}
\bibinfo{author}{\bibfnamefont{G.}~\bibnamefont{Arwas}},
  \bibinfo{author}{\bibfnamefont{D.}~\bibnamefont{Cohen}},
  \bibinfo{author}{\bibfnamefont{F.}~\bibnamefont{Hekking}}, \bibnamefont{and}
  \bibinfo{author}{\bibfnamefont{A.}~\bibnamefont{Minguzzi}},
  \bibinfo{journal}{Phys. Rev. A} \textbf{\bibinfo{volume}{96}},
  \bibinfo{pages}{063616} (\bibinfo{year}{2017}).

\bibitem[{\citenamefont{Arwas and Cohen}(2017)}]{Arwas-Cohen-PRB2017}
\bibinfo{author}{\bibfnamefont{G.}~\bibnamefont{Arwas}} \bibnamefont{and}
  \bibinfo{author}{\bibfnamefont{D.}~\bibnamefont{Cohen}},
  \bibinfo{journal}{Phys. Rev. B} \textbf{\bibinfo{volume}{95}},
  \bibinfo{pages}{054505} (\bibinfo{year}{2017}).

\bibitem[{\citenamefont{Arwas and Cohen}(2019)}]{Arwas-Cohen-PRA2019}
\bibinfo{author}{\bibfnamefont{G.}~\bibnamefont{Arwas}} \bibnamefont{and}
  \bibinfo{author}{\bibfnamefont{D.}~\bibnamefont{Cohen}},
  \bibinfo{journal}{Phys. Rev. A} \textbf{\bibinfo{volume}{99}},
  \bibinfo{pages}{023625} (\bibinfo{year}{2019}).

\bibitem[{\citenamefont{Eiermann et~al.}(2004)\citenamefont{Eiermann, Anker,
  Albiez, Taglieber, Treutlein, Marzlin, and
  Oberthaler}}]{Oberthaler-gap-soliton}
\bibinfo{author}{\bibfnamefont{B.}~\bibnamefont{Eiermann}},
  \bibinfo{author}{\bibfnamefont{T.}~\bibnamefont{Anker}},
  \bibinfo{author}{\bibfnamefont{M.}~\bibnamefont{Albiez}},
  \bibinfo{author}{\bibfnamefont{M.}~\bibnamefont{Taglieber}},
  \bibinfo{author}{\bibfnamefont{P.}~\bibnamefont{Treutlein}},
  \bibinfo{author}{\bibfnamefont{K.-P.} \bibnamefont{Marzlin}},
  \bibnamefont{and} \bibinfo{author}{\bibfnamefont{M.~K.}
  \bibnamefont{Oberthaler}}, \bibinfo{journal}{Phys. Rev. Lett.}
  \textbf{\bibinfo{volume}{92}}, \bibinfo{pages}{230401}
  (\bibinfo{year}{2004}).

\bibitem[{\citenamefont{Anker et~al.}(2005)\citenamefont{Anker, Albiez, Gati,
  Hunsmann, Eiermann, Trombettoni, and Oberthaler}}]{Oberthaler-selftrapped}
\bibinfo{author}{\bibfnamefont{T.}~\bibnamefont{Anker}},
  \bibinfo{author}{\bibfnamefont{M.}~\bibnamefont{Albiez}},
  \bibinfo{author}{\bibfnamefont{R.}~\bibnamefont{Gati}},
  \bibinfo{author}{\bibfnamefont{S.}~\bibnamefont{Hunsmann}},
  \bibinfo{author}{\bibfnamefont{B.}~\bibnamefont{Eiermann}},
  \bibinfo{author}{\bibfnamefont{A.}~\bibnamefont{Trombettoni}},
  \bibnamefont{and} \bibinfo{author}{\bibfnamefont{M.~K.}
  \bibnamefont{Oberthaler}}, \bibinfo{journal}{Phys. Rev. Lett.}
  \textbf{\bibinfo{volume}{94}}, \bibinfo{pages}{020403}
  (\bibinfo{year}{2005}).

\bibitem[{\citenamefont{Kartashov et~al.}(2011)\citenamefont{Kartashov,
  Malomed, and Torner}}]{RMP-solitons}
\bibinfo{author}{\bibfnamefont{Y.~V.} \bibnamefont{Kartashov}},
  \bibinfo{author}{\bibfnamefont{B.~A.} \bibnamefont{Malomed}},
  \bibnamefont{and} \bibinfo{author}{\bibfnamefont{L.}~\bibnamefont{Torner}},
  \bibinfo{journal}{Rev. Mod. Phys.} \textbf{\bibinfo{volume}{83}},
  \bibinfo{pages}{247} (\bibinfo{year}{2011}).

\bibitem[{\citenamefont{Bronski
  et~al.}(2001{\natexlab{a}})\citenamefont{Bronski, Carr, Deconinck, and
  Kutz}}]{Bronski}
\bibinfo{author}{\bibfnamefont{J.~C.} \bibnamefont{Bronski}},
  \bibinfo{author}{\bibfnamefont{L.~D.} \bibnamefont{Carr}},
  \bibinfo{author}{\bibfnamefont{B.}~\bibnamefont{Deconinck}},
  \bibnamefont{and} \bibinfo{author}{\bibfnamefont{J.~N.} \bibnamefont{Kutz}},
  \bibinfo{journal}{Phys. Rev. Lett.} \textbf{\bibinfo{volume}{86}},
  \bibinfo{pages}{1402} (\bibinfo{year}{2001}{\natexlab{a}}).

\bibitem[{\citenamefont{Wu and Niu}(2001)}]{Wu_Niu}
\bibinfo{author}{\bibfnamefont{B.}~\bibnamefont{Wu}} \bibnamefont{and}
  \bibinfo{author}{\bibfnamefont{Q.}~\bibnamefont{Niu}},
  \bibinfo{journal}{Phys. Rev. A} \textbf{\bibinfo{volume}{64}},
  \bibinfo{pages}{061603} (\bibinfo{year}{2001}).

\bibitem[{\citenamefont{Konotop and Salerno}(2002)}]{Konotop_Salerno}
\bibinfo{author}{\bibfnamefont{V.~V.} \bibnamefont{Konotop}} \bibnamefont{and}
  \bibinfo{author}{\bibfnamefont{M.}~\bibnamefont{Salerno}},
  \bibinfo{journal}{Phys. Rev. A} \textbf{\bibinfo{volume}{65}},
  \bibinfo{pages}{021602} (\bibinfo{year}{2002}).

\bibitem[{\citenamefont{Diakonov et~al.}(2002)\citenamefont{Diakonov, Jensen,
  Pethick, and Smith}}]{Pethick_Smith_PRA}
\bibinfo{author}{\bibfnamefont{D.}~\bibnamefont{Diakonov}},
  \bibinfo{author}{\bibfnamefont{L.~M.} \bibnamefont{Jensen}},
  \bibinfo{author}{\bibfnamefont{C.~J.} \bibnamefont{Pethick}},
  \bibnamefont{and} \bibinfo{author}{\bibfnamefont{H.}~\bibnamefont{Smith}},
  \bibinfo{journal}{Phys. Rev. A} \textbf{\bibinfo{volume}{66}},
  \bibinfo{pages}{013604} (\bibinfo{year}{2002}).

\bibitem[{\citenamefont{Machholm et~al.}(2003)\citenamefont{Machholm, Pethick,
  and Smith}}]{smith03}
\bibinfo{author}{\bibfnamefont{M.}~\bibnamefont{Machholm}},
  \bibinfo{author}{\bibfnamefont{C.~J.} \bibnamefont{Pethick}},
  \bibnamefont{and} \bibinfo{author}{\bibfnamefont{H.}~\bibnamefont{Smith}},
  \bibinfo{journal}{Phys. Rev. A} \textbf{\bibinfo{volume}{67}},
  \bibinfo{pages}{053613} (\bibinfo{year}{2003}).

\bibitem[{\citenamefont{Machholm et~al.}(2004)\citenamefont{Machholm, Nicolin,
  Pethick, and Smith}}]{smith04}
\bibinfo{author}{\bibfnamefont{M.}~\bibnamefont{Machholm}},
  \bibinfo{author}{\bibfnamefont{A.}~\bibnamefont{Nicolin}},
  \bibinfo{author}{\bibfnamefont{C.~J.} \bibnamefont{Pethick}},
  \bibnamefont{and} \bibinfo{author}{\bibfnamefont{H.}~\bibnamefont{Smith}},
  \bibinfo{journal}{Phys. Rev. A} \textbf{\bibinfo{volume}{69}},
  \bibinfo{pages}{043604} (\bibinfo{year}{2004}).

\bibitem[{\citenamefont{Seaman et~al.}(2005{\natexlab{a}})\citenamefont{Seaman,
  Carr, and Holland}}]{Holland-Kronig-Penney}
\bibinfo{author}{\bibfnamefont{B.~T.} \bibnamefont{Seaman}},
  \bibinfo{author}{\bibfnamefont{L.~D.} \bibnamefont{Carr}}, \bibnamefont{and}
  \bibinfo{author}{\bibfnamefont{M.~J.} \bibnamefont{Holland}},
  \bibinfo{journal}{Phys. Rev. A} \textbf{\bibinfo{volume}{71}},
  \bibinfo{pages}{033622} (\bibinfo{year}{2005}{\natexlab{a}}).

\bibitem[{\citenamefont{Seaman et~al.}(2005{\natexlab{b}})\citenamefont{Seaman,
  Carr, and Holland}}]{holland05}
\bibinfo{author}{\bibfnamefont{B.~T.} \bibnamefont{Seaman}},
  \bibinfo{author}{\bibfnamefont{L.~D.} \bibnamefont{Carr}}, \bibnamefont{and}
  \bibinfo{author}{\bibfnamefont{M.~J.} \bibnamefont{Holland}},
  \bibinfo{journal}{Phys. Rev. A} \textbf{\bibinfo{volume}{72}},
  \bibinfo{pages}{033602} (\bibinfo{year}{2005}{\natexlab{b}}).

\bibitem[{\citenamefont{Mu\~noz Mateo et~al.}(2019)\citenamefont{Mu\~noz Mateo,
  Delgado, Guilleumas, Mayol, and Brand}}]{Guilleumas-nonlinear_ring}
\bibinfo{author}{\bibfnamefont{A.}~\bibnamefont{Mu\~noz Mateo}},
  \bibinfo{author}{\bibfnamefont{V.}~\bibnamefont{Delgado}},
  \bibinfo{author}{\bibfnamefont{M.}~\bibnamefont{Guilleumas}},
  \bibinfo{author}{\bibfnamefont{R.}~\bibnamefont{Mayol}}, \bibnamefont{and}
  \bibinfo{author}{\bibfnamefont{J.}~\bibnamefont{Brand}},
  \bibinfo{journal}{Phys. Rev. A} \textbf{\bibinfo{volume}{99}},
  \bibinfo{pages}{023630} (\bibinfo{year}{2019}).

\bibitem[{\citenamefont{Carr et~al.}(2000{\natexlab{a}})\citenamefont{Carr,
  Clark, and Reinhardt}}]{carr00-a}
\bibinfo{author}{\bibfnamefont{L.~D.} \bibnamefont{Carr}},
  \bibinfo{author}{\bibfnamefont{C.~W.} \bibnamefont{Clark}}, \bibnamefont{and}
  \bibinfo{author}{\bibfnamefont{W.~P.} \bibnamefont{Reinhardt}},
  \bibinfo{journal}{Phys. Rev. A} \textbf{\bibinfo{volume}{62}},
  \bibinfo{pages}{063610} (\bibinfo{year}{2000}{\natexlab{a}}).

\bibitem[{\citenamefont{Carr et~al.}(2000{\natexlab{b}})\citenamefont{Carr,
  Clark, and Reinhardt}}]{carr00-b}
\bibinfo{author}{\bibfnamefont{L.~D.} \bibnamefont{Carr}},
  \bibinfo{author}{\bibfnamefont{C.~W.} \bibnamefont{Clark}}, \bibnamefont{and}
  \bibinfo{author}{\bibfnamefont{W.~P.} \bibnamefont{Reinhardt}},
  \bibinfo{journal}{Phys. Rev. A} \textbf{\bibinfo{volume}{62}},
  \bibinfo{pages}{063611} (\bibinfo{year}{2000}{\natexlab{b}}).

\bibitem[{\citenamefont{P\'erez-Obiol and Cheon}(2020)}]{Obiol-Cheon}
\bibinfo{author}{\bibfnamefont{A.}~\bibnamefont{P\'erez-Obiol}}
  \bibnamefont{and} \bibinfo{author}{\bibfnamefont{T.}~\bibnamefont{Cheon}},
  \bibinfo{journal}{Phys. Rev. E} \textbf{\bibinfo{volume}{101}},
  \bibinfo{pages}{022212} (\bibinfo{year}{2020}).

\bibitem[{\citenamefont{Abramowitz and Stegun}(1964)}]{abramowitz1964handbook}
\bibinfo{author}{\bibfnamefont{M.}~\bibnamefont{Abramowitz}} \bibnamefont{and}
  \bibinfo{author}{\bibfnamefont{I.~A.} \bibnamefont{Stegun}},
  \emph{\bibinfo{title}{Handbook of mathematical functions with formulas,
  graphs, and mathematical tables}}, vol.~\bibinfo{volume}{55}
  (\bibinfo{publisher}{US Government printing office}, \bibinfo{year}{1964}).

\bibitem[{\citenamefont{Byrd and Friedman}(2013)}]{byrd13book}
\bibinfo{author}{\bibfnamefont{P.~F.} \bibnamefont{Byrd}} \bibnamefont{and}
  \bibinfo{author}{\bibfnamefont{M.~D.} \bibnamefont{Friedman}},
  \emph{\bibinfo{title}{Handbook of elliptic integrals for engineers and
  physicists}}, vol.~\bibinfo{volume}{67} (\bibinfo{publisher}{Springer},
  \bibinfo{year}{2013}).

\bibitem[{\citenamefont{Kanamoto et~al.}(2008)\citenamefont{Kanamoto, Carr, and
  Ueda}}]{Carr_Ueda_PRL}
\bibinfo{author}{\bibfnamefont{R.}~\bibnamefont{Kanamoto}},
  \bibinfo{author}{\bibfnamefont{L.~D.} \bibnamefont{Carr}}, \bibnamefont{and}
  \bibinfo{author}{\bibfnamefont{M.}~\bibnamefont{Ueda}},
  \bibinfo{journal}{Phys. Rev. Lett.} \textbf{\bibinfo{volume}{100}},
  \bibinfo{pages}{060401} (\bibinfo{year}{2008}).

\bibitem[{\citenamefont{Wu and Niu}(2000)}]{Wu_Niu-landau-zener}
\bibinfo{author}{\bibfnamefont{B.}~\bibnamefont{Wu}} \bibnamefont{and}
  \bibinfo{author}{\bibfnamefont{Q.}~\bibnamefont{Niu}},
  \bibinfo{journal}{Phys. Rev. A} \textbf{\bibinfo{volume}{61}},
  \bibinfo{pages}{023402} (\bibinfo{year}{2000}).

\bibitem[{\citenamefont{Louis et~al.}(2003)\citenamefont{Louis, Ostrovskaya,
  Savage, and Kivshar}}]{louis03}
\bibinfo{author}{\bibfnamefont{P.~J.~Y.} \bibnamefont{Louis}},
  \bibinfo{author}{\bibfnamefont{E.~A.} \bibnamefont{Ostrovskaya}},
  \bibinfo{author}{\bibfnamefont{C.~M.} \bibnamefont{Savage}},
  \bibnamefont{and} \bibinfo{author}{\bibfnamefont{Y.~S.}
  \bibnamefont{Kivshar}}, \bibinfo{journal}{Phys. Rev. A}
  \textbf{\bibinfo{volume}{67}}, \bibinfo{pages}{013602}
  (\bibinfo{year}{2003}).

\bibitem[{\citenamefont{Kanamoto et~al.}(2009)\citenamefont{Kanamoto, Carr, and
  Ueda}}]{carr09}
\bibinfo{author}{\bibfnamefont{R.}~\bibnamefont{Kanamoto}},
  \bibinfo{author}{\bibfnamefont{L.~D.} \bibnamefont{Carr}}, \bibnamefont{and}
  \bibinfo{author}{\bibfnamefont{M.}~\bibnamefont{Ueda}},
  \bibinfo{journal}{Phys. Rev. A} \textbf{\bibinfo{volume}{79}},
  \bibinfo{pages}{063616} (\bibinfo{year}{2009}).

\bibitem[{\citenamefont{Bronski
  et~al.}(2001{\natexlab{b}})\citenamefont{Bronski, Carr, Deconinck, and
  Kutz}}]{bronski01}
\bibinfo{author}{\bibfnamefont{J.~C.} \bibnamefont{Bronski}},
  \bibinfo{author}{\bibfnamefont{L.~D.} \bibnamefont{Carr}},
  \bibinfo{author}{\bibfnamefont{B.}~\bibnamefont{Deconinck}},
  \bibnamefont{and} \bibinfo{author}{\bibfnamefont{J.~N.} \bibnamefont{Kutz}},
  \bibinfo{journal}{Phys. Rev. Lett.} \textbf{\bibinfo{volume}{86}},
  \bibinfo{pages}{1402} (\bibinfo{year}{2001}{\natexlab{b}}).

\bibitem[{\citenamefont{Wu et~al.}(2002)\citenamefont{Wu, Diener, and
  Niu}}]{wu02}
\bibinfo{author}{\bibfnamefont{B.}~\bibnamefont{Wu}},
  \bibinfo{author}{\bibfnamefont{R.~B.} \bibnamefont{Diener}},
  \bibnamefont{and} \bibinfo{author}{\bibfnamefont{Q.}~\bibnamefont{Niu}},
  \bibinfo{journal}{Phys. Rev. A} \textbf{\bibinfo{volume}{65}},
  \bibinfo{pages}{025601} (\bibinfo{year}{2002}).

\end{thebibliography}

\end{document}